\documentclass[12pt]{article}
\usepackage{amsfonts}
\usepackage{amsmath}
\usepackage{amssymb}
\usepackage{amsthm}
\usepackage{graphicx}
\usepackage{threeparttable}
\usepackage{threeparttablex}
\usepackage{booktabs}
\usepackage[font=bf,justification=centering]{caption}
\usepackage[top=1in, bottom=1in, left=1in, right=1in]{geometry}
\usepackage[onehalfspacing]{setspace}
\usepackage[backend=biber,bibstyle=mybibstyle,citestyle=authoryear-comp,uniquename=false,uniquelist=false]{biblatex}
\usepackage[toc,page]{appendix}
\usepackage[hidelinks,bookmarks]{hyperref}

\hypersetup{pdfstartview={XYZ null null 1.00}} % 100% zoom when opening PDF
\allowdisplaybreaks
\csname @openrightfalse\endcsname % Suppress page break after appendix title

\title{Monetary Policy and Firm Dynamics\thanks{I thank Morten Ravn, Vincent Sterk and seminar participants at UCL for useful feedback. The views in this paper do not reflect those of the Reserve Bank of Australia, whose financial support I gratefully acknowledge.}}
\date{\today}
\author{Matthew Read\thanks{University College London, Department of Economics, 30 Gordon Street, London, WC1H 0AX, United Kingdom. Email: matthew.read.16@ucl.ac.uk}}

\begin{document}

\maketitle

\begin{abstract}
    \noindent Do firm dynamics matter for the transmission of monetary policy? Empirically, the startup rate declines following a monetary contraction, while the exit rate increases, both of which reduce aggregate employment. I present a model that combines firm dynamics in the spirit of \textcite{Hopenhayn_1992} with New-Keynesian frictions and calibrate it to match cross-sectional evidence. The model can qualitatively account for the responses of entry and exit rates to a monetary policy shock. However, the responses of macroeconomic variables closely resemble those in a representative-firm model. I discuss the equilibrium forces underlying this approximate equivalence, and what may overturn this result.
\end{abstract}



\newpage

\section{Introduction}
\label{sec:introduction}

\begin{figure}[h]
    \center
    \caption{US Establishment Birth and Death Rates}
    \includegraphics[scale=0.75]{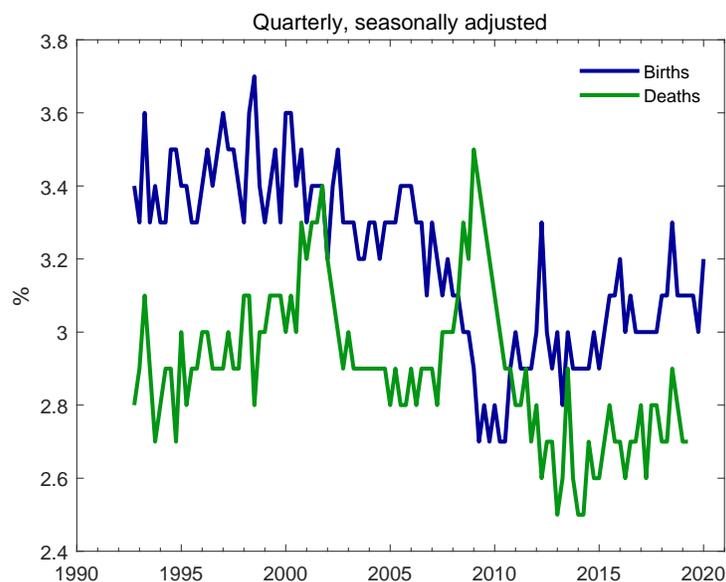}
    \label{fig:Establishments}
    \vspace{10 pt}\footnotesize \parbox[t]{0.65 in}{Source:}\parbox[t]{5 in}{US Bureau of Labor Statistics Business Employment Dynamics}
\end{figure}

Substantial rates of firm entry and exit are distinct features of the US economy; for example, quarterly entry and exit rates have each averaged around 3~per cent since 1990 and have undergone clear cyclical fluctuations (Figure~\ref{fig:Establishments}).\footnote{I refer to `establishments' and `firms' interchangeably, although in reality a firm may operate multiple establishments. Data used in the paper are for establishments rather than firms. I use `firm dynamics' to refer to the endogenous entry and exit of heterogeneous firms and associated changes in the productivity distribution.} Additionally, these margins contribute substantively to flows into and out of employment. In this paper, I provide empirical evidence suggesting that the entry and exit behaviour of firms may play an important role in monetary policy transmission. Overall, my results suggest that the primary extensive-margin response of the economy to a contractionary monetary policy shock is via an increase in firm exit. Moreover, I document that changes in job creation and destruction due to changes in entry and exit are sizeable relative to the change in aggregate employment following the shock. I explore the ability of a New Keynesian firm-dynamics model to explain the responses of firm entry and exit and investigate how firm dynamics alter the effects of monetary policy relative to a representative-firm benchmark.

The empirical evidence in this paper is based on a proxy structural vector autoregression (SVAR) in which the monetary policy shock is identified using high-frequency monetary policy surprises (as in \textcite{Gertler_Karadi_2015}). Estimates from the proxy SVAR suggest that the primary extensive-margin response of the economy following a contractionary monetary policy shock occurs through an increase in firm exit. The response of the entry rate is imprecisely estimated, but the point estimate indicates that the entry rate declines by a smaller magnitude than the change in the exit rate. Back-of-the-envelope calculations suggest that the extensive margin may contribute substantially to the change in aggregate employment following the shock; for example, the increase in job destruction due to higher firm exit in the four years after the shock is equivalent to almost 30~per cent of the change in employment, while the decline in job creation due to lower firm entry is equivalent to almost 20~per cent of the change in employment. The responses of firm entry and exit therefore result in a reduction in employment equivalent to almost half of the total decline in employment over this horizon.

To explore the implications of firm entry and exit for monetary policy transmission, I develop a New Keynesian firm-dynamics model that qualitatively matches the empirical responses of entry and exit to a monetary policy shock. The model incorporates heterogeneous firms in the spirit of \textcite{Hopenhayn_1992} and \textcite{Hopenhayn_Rogerson_1993} into an otherwise textbook New Keynesian model (e.g., \cite{Gali_2008}). In the model, heterogeneous firms experience idiosyncratic productivity shocks and face fixed entry and operating costs, which induce them to endogenously enter and exit production. These firms sell their output to intermediate goods producers, who set prices subject to a nominal rigidity. This structure for production generates a New Keynesian Phillips Curve linking the behaviour of the heterogeneous firms to aggregate inflation dynamics through the relative price of the heterogeneous firms' output, which is the real marginal cost faced by intermediate goods producers. To focus on the role of entry and exit in the transmission of monetary policy, the model abstracts from firm-level frictions. I calibrate the model to match features of the US economy and explore the model's predictions about the effects of a monetary policy shock.

In the model, a contractionary monetary policy shock causes an increase in the exit rate on impact and a smaller-magnitude decline in the entry rate, which is qualitatively consistent with the empirical evidence. The shock affects entry and exit rates through its effects on the three relative prices that enter the heterogeneous firms' decision problems. Focusing on the response of the exit rate, an increase in the real interest rate means that firms place less value on continuing to operate when deciding whether to pay the fixed operating cost, which makes them more likely to exit. The shock also sets off declines in the relative price of the heterogeneous firms' output (making exit more likely) and the real wage (making exit less likely). The effects of the declines in these two relative prices largely offset one another. Consequently, the overall effect of the shock on the exit rate is predominately driven by the direct effect of the higher real interest rate.

The responses of entry and exit rates to the monetary policy shock induce an endogenous change in aggregate total factor productivity (TFP), which is a channel of monetary policy transmission that is absent in the standard New Keynesian model.\footnote{\textcite{Moran_Queralto_2018} develop a New Keynesian model in which TFP is endogenously determined by the decisions of firms to invest in research and development.} The decrease in the entry rate and the increase in the exit rate reduce the total measure of operating firms, which pushes down aggregate TFP (due to decreasing returns to scale). At the same time, the entry and exit rates of less-productive firms are more sensitive to the monetary policy shock than those of more-productive firms, so the changes in entry and exit rates result in an upwards shift in the productivity distribution, which pushes up aggregate TFP. The former effect more than offsets the latter, so aggregate TFP falls. Changes in entry and exit rates of similar magnitudes to those estimated in the data result in an extremely persistent, albeit quantitatively small, change in aggregate TFP.

Although changes in entry and exit rates serve to endogenously amplify and propagate the monetary policy shock, impulse responses in the firm-dynamics model are almost identical to those in a comparable representative-firm model. This result is partly due to the quantitatively small responses of entry and exit rates to the shock, but also reflects the small size of firms that are induced to exit (or not enter) due to the shock. Together, these factors mean that the additional change in labour demand due to the extensive margin of adjustment is insubstantial relative to the change in labour demand due to changes in employment along the intensive margin. Furthermore, the role of entry and exit in amplifying the effects of the shock is dampened by changes in prices in general equilibrium. As a result, there is little additional amplification of the shock due to the entry and exit margins. I show that this result is robust to alternative calibrations and assumptions about the structure of the model. Finally, I discuss how the presence of firm-level frictions may increase the roles of entry and exit in the transmission of monetary policy.

\bigskip

\noindent\textbf{Relation to literature.} This work relates to an existing literature that uses New Keynesian models to explore the roles of entry and exit in transmitting monetary policy shocks. However, entry and exit in this literature are typically interpretable as the introduction or obsolescence of different varieties of consumption goods produced by monopolistically competitive firms that are otherwise homogeneous (e.g., \textcite{Lewis_2006}, \textcite{Bilbiie_Ghironi_Melitz_2007}, \textcite{Bergin_Corsetti_2008} and \textcite{Bilbiee_Fujiwara_Ghironi_2014}). Importantly, these models do not include any meaningful heterogeneity or dynamics at the firm level, which are features of firms' behaviour that have been well-documented using micro data (e.g., \cite{Foster_Haltiwanger_Krizan_2001}). These models therefore cannot speak to the role of endogenous and persistent changes in the distribution of firms in amplifying and propagating monetary policy shocks. In contrast, firms in my model are heterogeneous with respect to their productivity, as in \textcite{Hopenhayn_1992} and \textcite{Hopenhayn_Rogerson_1993}. The model therefore allows the potential for distributional dynamics to play a role in monetary policy transmission.

Other papers incorporate heterogeneous firms into New Keynesian models, but do not consider the roles of entry and exit in transmitting monetary policy shocks. In particular, \textcite{Ottonello_Winberry_2020} develop a heterogeneous-firm New Keynesian model to explore the investment channel of monetary policy. Their model uses a similar structure for production to the model in this paper, in the sense that the source of the nominal rigidity is separated from the heterogeneous firms, but it also includes physical capital and financial frictions. Firms in their model may endogenously default, but the entry process is exogenous and the total measure of operating firms is fixed. \textcite{Adam_Weber_2019} develop a New Keynesian model in which heterogeneous firms make price-setting decisions subject to a nominal rigidity, but entry and exit in their model is exogenous. By abstracting from firm-level frictions, I provide a useful benchmark against which richer models can be compared. To the best of my knowledge, this is the first paper to document the (near) irrelevance of firm dynamics in an otherwise-standard New Keynesian framework.

More generally, this paper relates to the literature exploring how firm dynamics affect the transmission of macroeconomic shocks. For example, \textcite{Samaniego_2008}, \textcite{Clementi_Palazzo_2016} and \textcite{Lee_Mukoyama_2018} consider the effects of aggregate technology shocks in models with heterogeneous firms and endogenous entry and exit, but without nominal rigidities. My finding that firm dynamics are essentially irrelevant for the transmission of monetary policy is similar in spirit to the results in \textcite{Thomas_2002} and \textcite{Khan_Thomas_2008}, who find that lumpy investment is quantitatively irrelevant in general-equilibrium business-cycle models (without nominal rigidities). My findings also resonate with the `near-aggregation' result in the heterogeneous-household model of \textcite{Krusell_Smith_1998}.

My empirical results also contribute to a growing body of evidence linking monetary policy to changes in firms' entry and exit behaviour. \textcite{Bergin_Corsetti_2008}, \textcite{Lewis_Poilly_2012} and \textcite{Uuskula_2016} document that different measures of firm entry and exit respond to monetary policy shocks using SVARs in which the monetary policy shock is identified by assuming that variables such as output and prices do not respond contemporaneously to the shock. In contrast, I estimate the effects of a monetary policy shock on firm entry and exit using high-frequency monetary policy surprises in a proxy SVAR, as in \textcite{Gertler_Karadi_2015}. The advantage of this approach is that it allows all variables in the system to respond contemporaneously to the monetary policy shock and thus does not require questionable zero restrictions on the contemporaneous relationships among the variables in the VAR. Additionally, I provide new estimates that quantify the importance of firm entry and exit in driving changes in employment following a monetary policy shock.

\bigskip

\noindent\textbf{Outline.} The remainder of the paper is structured as follows. Section~\ref{sec:empirical} presents new time-series evidence about the responses of firm entry and exit to monetary policy shocks. Section~\ref{sec:model} develops a New Keynesian firm-dynamics model. Section~\ref{sec:quantitative} describes the calibration of the model and explores how a monetary policy shock affects the economy in the firm-dynamics model relative to in a comparable representative-firm model. This section also documents the robustness of the model's predictions to changes in the calibration and under alternative assumptions about the model's structure. Section~\ref{sec:conclusion} concludes.\footnote{A full set of replication files are available at \url{https://sites.google.com/view/matthewread/}.}

\section{Evidence from a Proxy SVAR}
\label{sec:empirical}

In this section, I provide new evidence about the effects of US monetary policy on firm entry and exit using a proxy SVAR (e.g., \cite{Mertens_Ravn_2013}). I follow \textcite{Gertler_Karadi_2015} by using high-frequency surprises in three-month-ahead fed funds futures as a proxy for the monetary policy shock.\footnote{The series of monetary policy surprises is taken from \textcite{Jarocinski_Karadi_2020}, which is an updated version of the series constructed by \textcite{Gurkaynak_Sack_Swanson_2005}.} The motivation behind this proxy is that any change in fed funds futures prices in a narrow window around Federal Open Market Committee announcements should only reflect unexpected news about the path of the federal funds rate rate due to the monetary policy announcement. The surprises should therefore be correlated with the monetary policy shock and are plausibly exogenous with respect to other structural shocks, which is sufficient to point-identify impulse responses to the monetary policy shock.

The reduced-form VAR specification is a quarterly analogue of the VAR in \textcite{Gertler_Karadi_2015} that has been augmented with variables measuring firm entry and exit rates. As measures of activity and prices, I include real GDP and the GDP deflator (in logs). The VAR includes the one-year Treasury yield as a measure of the stance of monetary policy instead of the federal funds rate; this is to exploit variation in interest rates due to forward guidance, particularly during the period when the federal funds rate was constrained by the zero lower bound. The VAR also includes the excess bond premium from \textcite{Gilchrist_Zakrajsek_2012} to control for the systematic response of monetary policy to financial conditions.\footnote{The excess bond premium is the component of a measure of corporate credit spreads that is orthogonal to predicted default rates and which has been shown to contain predictive information about economic activity. I use the update of this series constructed by Favara \textit{et al.} (2016)\nocite{Favara_etal_2016}.} I add establishment birth and death rates from the US Bureau of Labor Statistics' Business Employment Dynamics (BED) dataset as measures of entry and exit, respectively.\footnote{An establishment is an economic unit producing goods or services, typically at one location, and undertaking (mainly) one activity. An establishment birth is an establishment reporting positive employment in the third month of the quarter and zero employment in the third month of the previous four quarters. An establishment death is identified as an establishment reporting zero employment in the third month of four consecutive quarters following a quarter with positive employment (the death is recorded in the first of the four quarters with zero employment). Establishment birth and death rates are measured as a percentage of the average number of establishments in the previous and current quarters. See \textcite{Sadeghi_2008} for further details.} The model includes four lags and a constant. The sample begins in the September quarter 1992, which reflects the availability of data on entry and exit, and ends in the December quarter 2016, which reflects the availability of the monetary policy surprises.

\begin{figure}[h]
    \center
    \caption{Impulse Responses to Monetary Policy Shock in Proxy SVAR}
    \vspace{5pt}
    \label{fig:PSVAR}
    \includegraphics[scale=1]{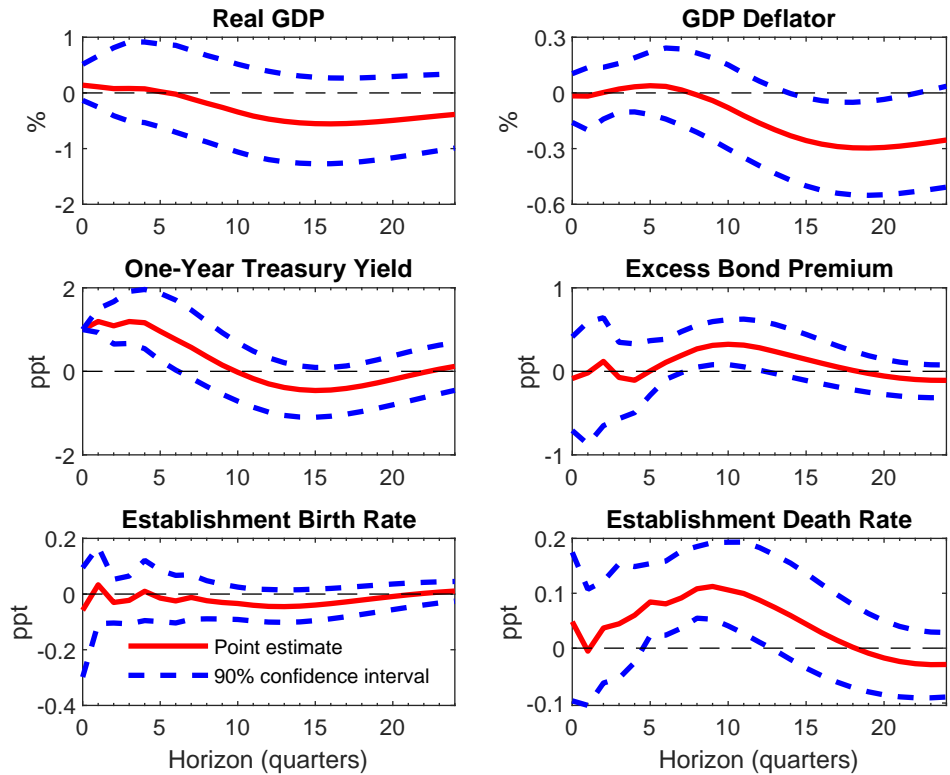}
    \footnotesize \parbox[t]{0.65 in}{Notes:}\parbox[t]{5 in}{Weak-instrument robust confidence intervals are computed using the approach of \textcite{Montiel-Olea_Stock_Watson_2020}.}
\end{figure}

Figure~\ref{fig:PSVAR} plots impulse responses to a monetary policy shock that increases the one-year Treasury yield by 100~basis points on impact. The monetary policy shock results in a persistent increase in the one-year Treasury yield and sluggish falls in output and prices. The peak decline in output is about 0.5~per cent and occurs after four~years, although the 90~per cent confidence intervals include zero at all horizons. After around five years, the GDP deflator has declined by about 0.3~per cent. These responses are consistent with predictions from standard macroeconomic theory. The excess bond premium increases, which is consistent with a tightening in financial conditions.

Following the shock, the quarterly establishment exit rate increases by about 10~basis points over the first two years and the confidence intervals exclude zero at horizons between one and three years. The entry rate declines by a maximum of about 4--5~basis points after around three years, although the confidence intervals include zero at all horizons.\footnote{These findings are broadly consistent with those in \textcite{Uuskula_2016}, who identifies the shock using a partial causal ordering with different variables in the VAR.} The results are qualitatively similar when including the number of entering and exiting establishments (in logs) rather than entry and exit rates. In response to a contractionary monetary policy shock, the number of exiting establishments increases by 1.4~per cent after about two years and the confidence intervals exclude zero at horizons between one and two years after the shock. The number of entering establishments decreases by 0.9~per cent after three years, although the confidence intervals include zero at all horizons.

To roughly quantify the contribution of firms' entry and exit behaviour to monetary policy transmission, I replace real GDP with the log of total nonfarm employment in the proxy SVAR that includes entry and exit in levels, and compute the approximate change in employment that is directly due to changes in the entry and exit margins following the shock. I assume that variables are initially at their December 2016 levels, and compute the contribution of the entry (exit) margin to the change in employment by cumulating the change in the number of entrants (exits) and multiplying by the average size of an entrant (exit) in the Business Dynamics Statistics (BDS).\footnote{The BDS are publicly released statistics aggregated from the US Census Bureau's Longitudinal Business Database. In this exercise, I assume that the initial level of employment is 145~million. The initial numbers of entering and exiting firms per quarter are, respectively, 239,000 and 217,000. Based on the 2016 release of the BDS, the average number of workers employed by entering firms is 8.2 and by exiting firms is 7.7.} This exercise assumes that firms that do not enter (or that exit) due to the shock are on average the same size as entrants (or exiting firms) unconditionally.

\begin{figure}[h]
    \center
    \caption{Employment Effects of Monetary Policy Shock in Proxy SVAR}
    \vspace{5pt}
    \label{fig:EmploymentEffects}
    \includegraphics[scale=1]{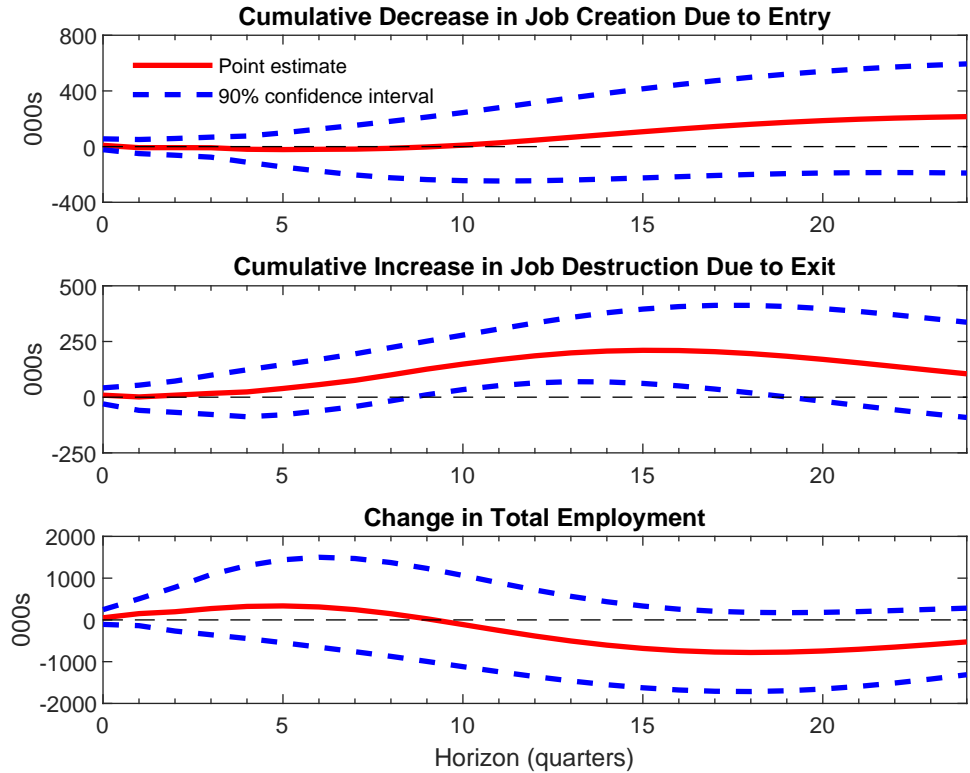}
    \footnotesize \parbox[t]{0.65 in}{Notes:}\parbox[t]{5 in}{Weak-instrument robust confidence intervals are computed using the approach of \textcite{Montiel-Olea_Stock_Watson_2020}.}
\end{figure}

Figure~\ref{fig:EmploymentEffects} summarises the results from this exercise. Four years after the shock, employment has declined by about 0.5~per cent, which is equivalent to about 740,000~workers. The increase in firm exit due to the shock means that an additional 210,000 jobs are destroyed in the four years following the shock, which is equivalent to almost 30~per cent of the total decline in employment. At the same time, the decrease in firm entry due to the shock means that around 125,000 fewer jobs are created in the four years following the shock, which is equivalent to a bit less than 20~per cent of the total decline in employment. Note, however, that the confidence intervals for the cumulative change in employment attributable to the entry margin include zero at all horizons. The point estimates suggest that the combined response along the extensive margin directly results in a reduction in employment that is equivalent to almost half of the total decline in employment four years after the shock.

Overall, the results from the proxy SVAR indicate that a contractionary monetary policy shock results in an increase in firm exit and a decrease in firm entry, although impulse responses of the measures of entry are not significantly different from zero at conventional levels of significance. The exit margin appears to be more responsive to a monetary policy shock than the entry margin, so the primary extensive-margin response of the economy to a monetary policy shock appears to be via firm exit rather than firm entry. Moreover, the estimates suggest that these margins may contribute substantively to the total change in employment following a monetary policy shock.

\section{A New Keynesian Firm-Dynamics Model}
\label{sec:model}

In this section, I develop a New Keynesian model with heterogeneous firms that endogenously enter and exit production. The model embeds the firm-dynamics model of \textcite{Hopenhayn_1992} and \textcite{Hopenhayn_Rogerson_1993} into a textbook New Keynesian model (e.g., \cite{Gali_2008}).

Time in the model is discrete (indexed by $t=0,1,...$) and all agents have rational expectations. For tractability, I follow \textcite{Ottonello_Winberry_2020} by segmenting production into different stages to separate the source of nominal rigidity from the heterogeneous firms' production decisions. `Production' firms competitively produce an undifferentiated good using labour as the sole input in a production function with decreasing returns to scale. These firms face idiosyncratic TFP shocks and fixed entry and operating costs, which induce them to enter and exit production. Intermediate goods producers purchase the undifferentiated good from production firms and use it to produce a differentiated good, which they sell in a monopolistically competitive market subject to a nominal rigidity. Final goods producers purchase the differentiated goods and bundle them into a final good. Households enjoy utility from leisure and consumption of the final good, and can save in a nominal bond that is in zero net supply. The central bank sets the interest rate on the nominal bond according to a Taylor rule. The government collects fixed costs from the production firms and distributes these to the household as a lump-sum transfer.

In the following subsections, I detail the decision problems faced by each type of agent and develop the equilibrium conditions.

\subsection{Households}
\label{subsec:households}

A representative household enjoys utility from consumption ($C_{t}$) and suffers disutility from supplying labour ($N_{t}$) at real wage rate $w_{t}$. The household has access to a nominal bond ($B_{t}$) paying gross nominal interest rate $R_{t}$. The household earns real dividends ($D_{t}$) from owning firms and receives real lump-sum transfers ($T_{t}$) from the government. The household's utility maximisation problem is
\begin{equation}
  \max_{\{C_{t},N_{t},B_{t}\}_{t=0}^{\infty}} \mathbb{E}_{0}\sum_{t=0}^{\infty}\beta^{t}\left(\frac{C_{t}^{1-\sigma}-1}{1-\sigma} - \kappa_{0} \frac{N_{t}^{1+\kappa_{1}}}{1+\kappa_{1}}\right)
\end{equation}
subject to the budget constraint
\begin{equation} \label{eq:budgetconstraint}
  C_{t} + \frac{B_{t}}{P_{t}} = R_{t-1}\frac{B_{t-1}}{P_{t}} + w_{t}N_{t} + D_{t} + T_{t},
\end{equation}
where $P_{t}$ is the price of the final good, $\beta$ is the subjective discount factor, $\kappa_{0}$ scales the disutility from supplying labour and $\kappa_{1}$ is the inverse of the Frisch elasticity of labour supply.

The first-order conditions for the problem imply the labour supply condition
\begin{equation}
    w_{t} = \kappa_{0}C_{t}^{\sigma}N_{t}^{\kappa_{1}}
\end{equation}
and the Euler equation
\begin{equation}
  1 = \mathbb{E}_{t}\left(\Lambda_{t,t+1}\frac{R_{t}}{\Pi_{t+1}}\right), \label{eq:euler_B}
\end{equation}
where $\Pi_{t+1} = P_{t+1}/P_{t}$ is the gross rate of inflation between periods $t$ and $t+1$, and $\Lambda_{t,t+k} = \beta^{k} (C_{t+k}/C_{t})^{-\sigma}$ is the representative household's $k$-period stochastic discount factor.

\subsection{Central bank and government}
\label{subsec:centralbank}

The central bank sets the nominal interest rate based on an inflation-targeting rule:
\begin{equation} \label{eq:taylorrule}
  \frac{R_{t}}{R} = \left(\frac{\Pi_{t}}{\Pi}\right)^{\phi}\exp(\varepsilon_{t}^{m}),
\end{equation}
where $R$ is the nominal interest rate in the stationary equilibrium and $\Pi$ is the inflation target. $\varepsilon_{t}^{m}$ is a monetary policy shock following the AR(1) process $\varepsilon_{t}^{m} = \rho_{m}\varepsilon_{t-1}^{m} + \eta_{t}$, where $\eta_{t}$ is white noise.\footnote{As in \textcite{Ottonello_Winberry_2020}, including persistence in the monetary policy shock is a simple way to induce persistence in the responses of variables to the shock. An alternative would be to include interest-rate smoothing in the Taylor rule.}

The only role of the government is to collect entry and operating costs paid by production firms and remit these to the household as a lump-sum transfer, which will mean that the fixed costs paid by production firms do not appear in any market clearing conditions. This facilitates comparison against the representative-firm model, in which there are no fixed costs.

\subsection{Production firms}
\label{subsec:productionfirms}

The structure for production of the undifferentiated good is similar to that in \textcite{Hopenhayn_Rogerson_1993}. There is a measure of heterogeneous production firms (indexed by $j$) producing an undifferentiated good according to the production function
\begin{equation}\label{eq:prodfn}
  y_{jt} = z_{jt}n_{jt}^{\nu},
\end{equation}
where $z_{jt} \in \mathbb{R}_+$ is idiosyncratic productivity, $n_{jt}$ is the firm's labour input, and $0 < \nu < 1$, so there are decreasing returns to scale. $z_{jt}$ follows an AR(1) process in logs:
\begin{equation}\label{eq:tfpprocess}
  \ln z_{jt} = a_{z}(1-\rho_{z}) + \rho_{z} \ln z_{j,t-1} + \varepsilon_{jt}^{z}, \quad \varepsilon_{jt}^{z} \overset{iid}{\sim} N(0,\sigma_{z}^{2}).
\end{equation}
The cumulative distribution function (CDF) of $z_{jt}$ conditional on $z_{j,t-1}$ is $F(.|z_{j,t-1})$.

As in \textcite{Clementi_Palazzo_2016}, production firms face a stochastic fixed operating cost, $c_{jt}$, denominated in units of the final good. The fixed cost is independently and identically distributed across firms and over time according to a lognormal distribution: $c_{jt} \overset{iid}{\sim} LN(\mu_{c},\sigma_{c})$ with CDF $G_{c}(.)$. At the beginning of each period, firms learn the current draw of $z_{jt}$, choose $n_{jt}$ and produce output. Firms then draw the fixed operating cost and must choose whether to pay this cost in the current period to continue operating in the next period or to shut down and earn zero profits thereafter.

There is also a fixed mass $M$ of potential entrants (indexed by $m$). At the beginning of each period, potential entrants draw $z_{mt}$ from a distribution with CDF $Q(.)$. Potential entrants who decide to begin operating incur a one-time fixed cost, $e_{mt}$, denominated in units of the final good. Unlike in \textcite{Hopenhayn_Rogerson_1993}, this cost is stochastic; in particular, $e_{mt} \overset{iid}{\sim} \times LN(\mu_{e},\sigma_{e})$ with CDF $G_{e}(.)$. Potential entrants draw the entry cost after drawing $z_{mt}$ but before deciding whether to enter. Entrants then operate in exactly the same way as an existing firm. Entry and operating costs are financed using equity.

The individual state variable for an operating firm is $z_{jt}$. The aggregate state variables are the distribution of incumbent firms over idiosyncratic productivity (described below) and the aggregate shock $\varepsilon_{t}^{m}$. Firms care about the aggregate state of the economy only to the extent that it affects prices, which they take as given. I summarise the aggregate state by including time, $t$, as an index in the firm's value function. For ease of notation, I henceforth drop firm and time subscripts from firm-specific variables and denote next-period values with a prime.

Production firms operate in a competitive output market and hire labour in a competitive spot market at real wage rate $w_{t}$. They discount the future using the representative household's stochastic discount factor. Let $V_{t}(z)$ be the expected discounted value of future real profits (in units of the final good) for an incumbent firm entering period $t$ with idiosyncratic productivity $z$ and who behaves optimally. Consider the decision problem of a firm that has just drawn fixed operating cost $c$ and must decide whether to pay this fixed cost and continue operating in the next period or to exit. The firm will only continue to operate if $\mathbb{E}_{t}\left(\Lambda_{t,t+1}V_{t+1}(z')\right) \geq c$. The left-hand side of this inequality implicitly depends on $z$ (because $z$ follows an AR(1) process) and is constant in $c$, while the right-hand side is increasing in $c$. There thus exists a threshold value of $c$ at each value of $z$, $c_{t}^{*}(z)=\mathbb{E}_{t}\left(\Lambda_{t,t+1}V_{t+1}(z')\right)$, below which the firm will continue to operate and above which it will exit.\footnote{If the fixed operating cost were not stochastic, as in \textcite{Hopenhayn_Rogerson_1993}, there would exist a threshold value of idiosyncratic productivity below which all firms would exit and above which all firms would choose to continue operating, which is inconsistent with observed patterns of firm exit.}

The Bellman equation for the firm's problem is
\begin{equation}\label{eq:bellman_existing1}
   V_{t}(z) = \max_{n\geq0}p_{t}zn^{\nu} - w_{t}n + \int\max\left\{\mathbb{E}_{t}\left(\Lambda_{t,t+1}V_{t+1}(z')\right)-c,0\right\}dG_{c}(c),
\end{equation}
where $p_{t}$ is the relative price of the undifferentiated good in units of the final good. The first-order condition for the problem implies the policy function for labour demand:
\begin{equation}\label{eq:decisionrule_N}
  n_{t}(z) = \left(\frac{w_{t}}{\nu p_{t} z}\right)^{\frac{1}{\nu-1}}.
\end{equation}
Using this policy function and the threshold for exit, the Bellman equation becomes
\begin{equation}\label{eq:bellman_existing2}
  V_{t}(z) = p_{t}zn_{t}(z)^{\nu} - w_{t}n_{t}(z) \\
  +\left[\mathbb{E}_{t}\left(\Lambda_{t,t+1}V_{t+1}(z')\right)-\mathbb{E}_{c}\left(c|c \leq c_{t}^{*}(z)\right)\right]G_{c}\left(c_{t}^{*}(z)\right).
\end{equation}
The term in square brackets is the expected discounted value of operating in time $t+1$ net of the expected fixed operating cost conditional on this cost being less than the expected discounted value of operating.\footnote{$\mathbb{E}_{c}\left(c|c \leq c_{t}^{*}(z)\right)$ is the expected value of a truncated lognormal random variable. If $x\sim LN(\mu,\sigma)$ and $\Phi(.)$ is the standard normal CDF, then
\begin{equation*}
  \mathbb{E}_{x}(x|x \leq k) = \exp\left(\mu + \frac{\sigma^2}{2}\right)\frac{\Phi\left(\frac{\ln k - \mu - \sigma^2}{\sigma}\right)}{\Phi\left(\frac{\ln k - \mu}{\sigma}\right)}.
\end{equation*}} This is multiplied by the probability that the fixed operating cost is less than the exit threshold. With the complementary probability, the fixed operating cost is greater than the exit threshold, in which case the firm will exit and will have zero continuation value.

Similarly, a potential entrant that has drawn idiosyncratic productivity $z$ and fixed entry cost $e$ will only begin operating if $V_{t}(z) \geq e$. There exists a threshold value of $e$ at each value of $z$, $e_{t}^{*}(z)=V_{t}(z)$, below which a potential entrant will choose to enter and above which it will not.

The measure of operating firms across idiosyncratic productivity in period $t$, $\mu_{t}(z)$, consists of firms that were operating in period $t-1$ that chose not to exit and new entrants in period $t$. For all Borel subsets $B \in \mathbb{R}_{+}$ of the idiosyncratic state space, this measure evolves according to
\begin{equation}\label{eq:mu_transition}
  \mu_{t+1}(B) = \int\int_{z' \in B}G_{c}\left(c_{t}^{*}(z)\right)dF(z'|z)d\mu_{t}(z) + M\int_{z' \in B} G_{e}\left(e_{t+1}^{*}(z')\right)dQ(z').
\end{equation}
This is the relevant measure for computing aggregate output, employment, etc.\footnote{The measure of operating firms over idiosyncratic productivity is not part of the aggregate state, since it is not predetermined; it depends on the entry decisions of firms, which in turn depend on prices in the current period. The aggregate state includes the measure of \textit{incumbent} firms, which is equal to the measure of operating firms less new entrants.
% $\tilde{\mu}_{t}$, which evolves according to
%\begin{equation*}
%  \tilde{\mu}_{t+1}(B) = \int \int_{z' \in B} G_{c}(c_{t}^{*}(z))dF(z'|z)d\tilde{\mu}_{t}(z) + M \int \int_{z' \in B} G_{c}(c_{t}^{*}(z))G_{e}(e_{t}^{*}(z))dF(z'|z)dQ(z).
%\end{equation*}
%The measure of operating firms is made up of incumbent firms and new entrants, so
} Aggregate output is $Y_{t} = \int zn_{t}(z)^{\nu}d\mu_{t}(z)$, aggregate employment is $N_{t} = \int n_{t}(z)d\mu_{t}(z)$ and aggregate real profit earned by the production firms is $\Omega_{t} = p_{t}Y_{t} - w_{t}N_{t} - T_{t}$, where $T_{t}$ are aggregate fixed operating and entry costs:
\begin{equation}\label{eq:transfer}
    T_{t} = \int G_{c}\left(c_{t}^{*}(z)\right)\mathbb{E}_{c}\left(c|c \leq c_{t}^{*}(z)\right)d\mu_{t}(z) + M\int G_{e}\left(e_{t}^{*}(z)\right)\mathbb{E}_{e}\left(e | e \leq e_{t}^{*}(z)\right)dQ(z).
\end{equation}

\subsection{Intermediate and final goods producers}
\label{subsec:intermediatefirms}

Since this part of the model is fairly standard, I describe it at a high level and relegate detail to Appendix~\ref{sec:equilibriumconditions}. There is a unit mass of intermediate goods producers, which purchase the undifferentiated good from the production firms and costlessly differentiate it. Intermediate goods producers sell their products to a representative final good producer in a monopolistically competitive market, taking the demand schedule of the final good producer as given. Intermediate goods producers face a quadratic price-adjustment cost -- paid in units of the final good -- as in \textcite{Rotemberg_1982}. The final good producer bundles intermediate goods into a final good according to a CES production function, which generates a downward-sloping demand schedule for the intermediate goods.

I consider symmetric equilibria where all intermediate goods firms face the same initial price, which implies that they will optimally choose the same price in each period. Intermediate firms' optimal price-setting behaviour is characterised by a (nonlinear) Phillips Curve:
\begin{equation} \label{eq:pricesetting}
  (1-\gamma) + \gamma p_{t} - \xi(\Pi_{t} - 1)\Pi_{t} = -\xi \mathbb{E}_{t}\left[\Lambda_{t,t+1}(\Pi_{t+1}-1)\Pi_{t+1}\frac{Y_{t+1}}{Y_{t}}\right].
\end{equation}
Real profit earned by the intermediate goods firms in a symmetric equilibrium is
\begin{equation}\label{eq:intermediategoodsfirmsprofit}
  \Upsilon_{t} = \left(1-p_{t}-\frac{\xi}{2}\left(\Pi_{t}-1\right)^{2}\right)Y_{t}.
\end{equation}
Aggregate dividends are then $D_{t} = \Omega_{t} + \Upsilon_{t}$.

\subsection{Equilibrium}
\label{subsec:equilibrium}

Given the law of motion for the exogenous state $\varepsilon_{t}^{m}$, the competitive equilibrium is a joint law of motion for production firms' policy functions and value function $\{n_{t}(z),c_{t}^{*}(z),e_{t}^{*}(z),V_{t}(z)\}_{z \in \mathbb{R}_+}$, the measure of operating firms over idiosyncratic productivity, $\{\mu_{t}(z)\}_{z \in \mathbb{R}_+}$, aggregate quantities $\{C_t,N_t,B_t,Y_t,D_t,T_t\}$ and (relative) prices $\{p_t,w_t,\Pi_t\}$ such that, for all $t$: the value function solves the production firms' Bellman equation for all $z \in \mathbb{R}_+$ with associated policy functions $n_{t}(z)$, $c_{t}^{*}(z)$ and $e_{t}^{*}(z)$; the Euler equation and labour supply condition are satisfied; the Taylor Rule and Phillips Curve are satisfied; $\mu_{t}(z)$ evolves according to the transition function in Equation~\ref{eq:mu_transition}; the profits of production firms and intermediate goods producers are paid out as dividends; fixed costs are remitted to the household according to Equation~\ref{eq:transfer}; and goods, labour and bond markets clear.\footnote{The labour market clearing condition is $N_{t} = \int n_{t}(z)d\mu_{t}(z)$, the bond market clearing condition is $B_{t} = 0$ (since bonds are in zero net supply) and the (final) goods market clearing condition is $Y_{t} = C_{t} + \frac{\xi}{2}(\Pi_{t}-1)^{2}Y_{t}$. Clearing of the markets for production and intermediate goods is implicit in the Phillips Curve.}

\subsection{Stationary Equilibrium}
\label{subsec:statequilibrium}

In a stationary equilibrium without aggregate risk, all aggregate quantities and relative prices are constant. The measure of firms across idiosyncratic productivity is also constant, although the position of individual firms will move around within the distribution, and firms will enter and exit.

I consider a stationary equilibrium with zero net inflation, so $\Pi = 1$. The Euler equation implies that $R = 1/\beta$ and intermediate firms' price-setting condition implies that $p = (\gamma-1)/\gamma$. Given $p$ and an arbitrary value of $w$, solving the production firms' problem yields the value function and the policy functions for labour demand, entry and exit. The stationary measure of firms is then a fixed point of
\begin{equation}
  \mu(B) = \int\int_{z' \in B}G_{c}\left(c^{*}(z)\right)dF(z'|z)d\mu(z) + M\int_{z' \in B} G_{e}\left(e^{*}(z')\right)dQ(z'),
\end{equation}
which can be used to compute aggregate output $Y = \int z n(z)^{\nu}d\mu(z)$ and labour demand $N = \int n(z)d\mu(z)$. Given the value of $w$ and imposing labour market clearing then implies a unique value of consumption $C$ from the labour supply condition. One could then check whether the final good market clears (i.e., $Y = C$). Since the choice of $w$ was arbitrary, clearing of the final good market pins down the real wage in the stationary equilibrium.\footnote{Under the calibration of the model described below, I have verified numerically that there is a unique stationary equilibrium; aggregate supply of the final good is decreasing in $w$ and, given that the labour market clears, aggregate demand for the final good is increasing in $w$.}

\section{Quantitative Results}
\label{sec:quantitative}

The goal of this section is to explore the ability of the model developed in Section~\ref{sec:model} to explain the responses of firm entry and exit following a monetary policy shock, and to examine how allowing for firm dynamics alters the effects of the shock relative to a representative-firm benchmark. To this end, I first calibrate the model so that features of its stationary equilibrium broadly match US data. I then solve for the model's equilibrium dynamics in the vicinity of the stationary equilibrium using first-order perturbation.

\subsection{Calibration}
\label{subsec:calibration}

The model period is one quarter. The target for the annual real interest rate is 4~per cent, so $\beta = 1.04^{-1/4}$. $\sigma = 1$, so there is log utility in consumption. $\kappa_{1} = 1$, so there is a unit Frisch elasticity. $\gamma = 6$, so intermediate goods producers set a markup of 20~per cent in the stationary equilibrium. $\phi = 1.5$, so the central bank moves the nominal interest rate more than one-for-one in response to deviations of inflation from target. These choices are within the range of values usually considered in the literature. $\nu$, which controls returns to scale, is 0.9, which is within the range of estimates in \textcite{Basu_Fernald_1997} and \textcite{Lee_2005}. $\xi$ is set to 50 to deliver a Phillips Curve slope of 0.1, as in \textcite{Ottonello_Winberry_2020}.

In a stationary equilibrium, the log of firm-level employment follows the AR(1) process
\begin{equation}\label{eq:lom_N}
  \ln n_{jt} = \frac{1-\rho_{z}}{\nu - 1}\ln \frac{\nu}{p a_z} + \rho_{z}\ln n_{j,t-1} + \left(\frac{1}{\nu -1}\right)\varepsilon_{jt}^{z}.
\end{equation}
Using annual data from the US Census Bureau's Longitudinal Business Database, \textcite{Pugsley_Sedlacek_Sterk_2020} estimate the parameters of different processes for firm-level employment -- including an AR(1) process -- by matching the observed cross-sectional autocovariance structure of employment. Based on a balanced panel of firms, they estimate an AR(1) persistence parameter of $\rho_{n} = 0.9771$ and an innovation standard deviation of $\sigma_{n} = 0.2676$.\footnote{See column (5) in Table B.2 in their online appendix.} The implied persistence parameter in the quarterly process for log productivity is $\rho_{z} = \rho_{n}^{1/4}$ and the implied innovation standard deviation is $\sigma_{z} = (\nu - 1)\sqrt{\sigma_{n}^{2}/\sum_{j=0}^{3}\rho_{z}^{2j}}$.

I discretise the process for idiosyncratic productivity using the method of \textcite{Rouwenhorst_1995} (as described in \textcite{Kopecky_Suen_2010}) with 50~evenly spaced points for $\ln z$ and solve the Bellman equation for production firms on this grid using value function iteration. The distribution from which potential entrants draw idiosyncratic productivity, $Q(z)$, is the same as the unconditional distribution faced by existing firms. The distribution of entry costs coincides with the distribution of fixed operating costs, so $\mu_e = \mu_c$, $\sigma_e = \sigma_c$ and $G_e(.) = G_c(.)$. The stationary measure of firms is obtained by iterating on its transition function to convergence.

I normalise the real wage $w$ to unity and find $M$ such that the final good market clears.\footnote{The measure of firms is homogeneous of degree one in $M$, and thus so is aggregate employment. Given targets for the average size of incumbent firms and the employment-to-population ratio, $M$ essentially adjusts so that the measure of firms reconciles these two moments (at $w = 1$).} I jointly calibrate the remaining parameters $(\mu_{c},\sigma_{c},a_{z},\kappa_{0})$ to target the annual exit rate (8.6~per cent), the average size of an incumbent firm (19.2~employees), the average size of exiting firms (7.7~employees), and the employment-to-population ratio (0.6). The first three targets are taken from the 2016 BDS, while the last is roughly equal to the 2016 employment-to-population ratio based on the Bureau of Labor Statistics' Current Population Survey.\footnote{I estimate the average size of an incumbent firm in the BDS by subtracting jobs created due to establishment births from total employment and dividing by the total number of establishments less the number of establishment births.} Table~\ref{tab:parameters} lists the parameter values under this calibration.

\begin{table}[h]
    \small
    \centering
    \setlength{\tabcolsep}{10pt}
    \begin{threeparttable}
        \captionsetup{justification=centering}
        \caption{Calibrated Parameters} \label{tab:parameters}
        \center
        \begin{tabular}{c c c c}
            \toprule
            {Parameter}  & {Value} & {Parameter}   & {Value}  \\ \midrule
$\beta$ &	0.99	& $\kappa_{0}$ &	2.083	\\
$\sigma$ &	1	& $\kappa_{1}$ &	1	\\
$\nu$ &	0.9	& $\phi$ &	1.5	\\
$\gamma$ &	6	& $\xi$ &	50	\\
$\mu_{c}$ &	--6.216	& $\sigma_{c}$ &	4.537	\\
$\sigma_{z}$ &	0.013	& $\rho_{z}$ &	0.994	\\
$M$ &	$7.483\times 10^{-4}$	& $a_{z}$ &	0.439	\\
            \bottomrule
        \end{tabular}
%        \begin{tablenotes}[flushleft]
%            \item Notes:
%        \end{tablenotes}
    \end{threeparttable}
\end{table}

\begin{figure}[h]
    \center
    \caption{Distribution of Firms by Size}
    \label{fig:firmDistribution}
    \includegraphics[scale=0.6]{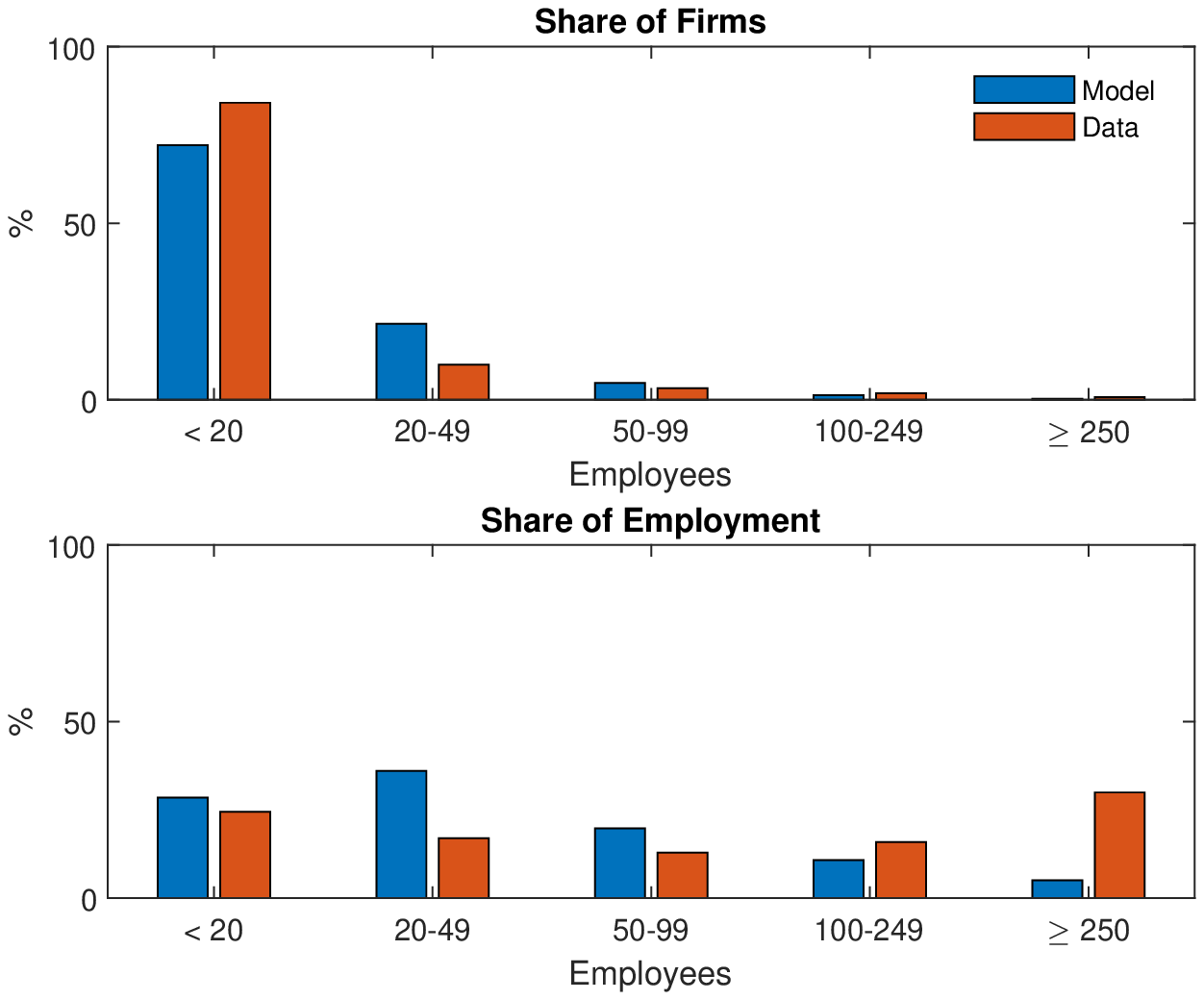} \hspace{10pt} \includegraphics[scale=0.6]{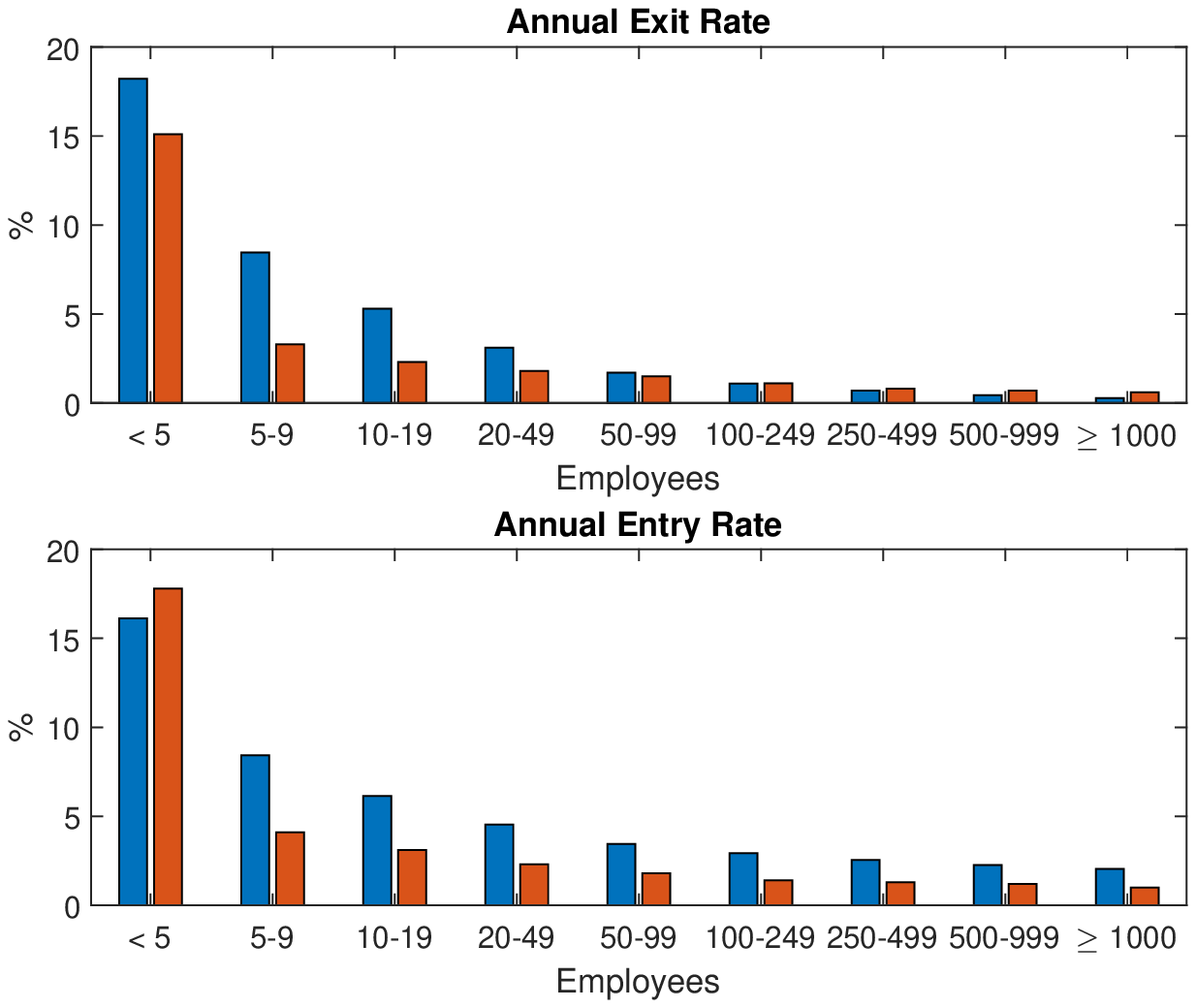} \\
    \footnotesize \parbox[t]{0.65 in}{Notes:}\parbox[t]{5 in}{Data are from 2016 BDS.}
\end{figure}

The targeted moments in the model are equal to the moments in the data. The model also appears to perform reasonably well in matching non-targeted features of the distribution of firms, particularly given the simplicity of the model and the small set of calibration targets. As in the data, the model predicts that there are more small firms than large firms, although large firms account for a smaller share of total employment in the model than in the data (Figure~\ref{fig:firmDistribution}, left panels). The profiles of entry and exit rates in the model and data are similar in the sense that entry and exit rates decline with firm size (Figure~\ref{fig:firmDistribution}, right panels). The model also closely matches the age distribution of firms and the share of employment by age in the data.

\begin{figure}[h]
    \center
    \caption{Entry and Exit Probabilities by Idiosyncratic Productivity}
    \label{fig:partialEqResponses}
    \includegraphics[scale=0.6]{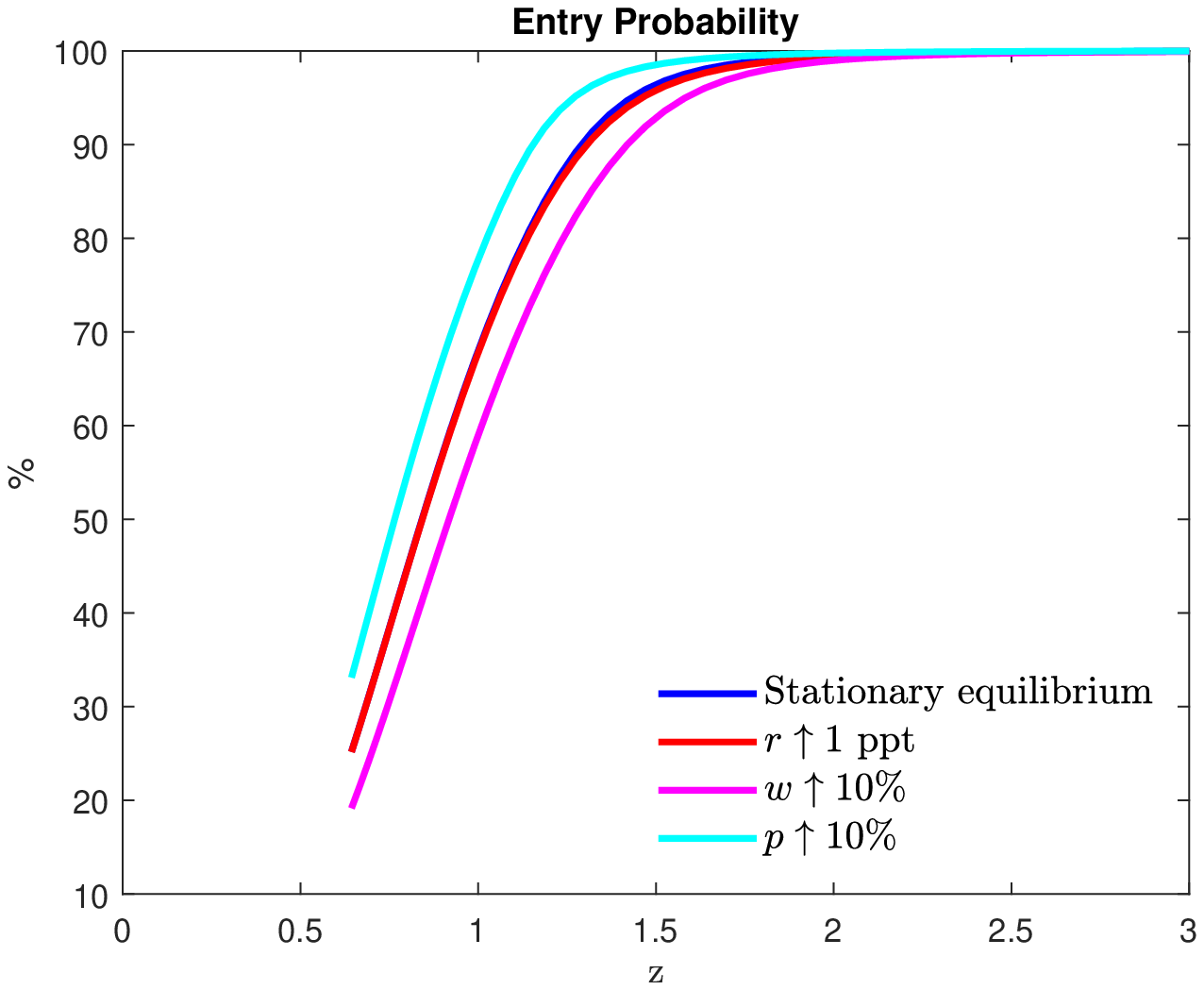} \hspace{10pt} \includegraphics[scale=0.6]{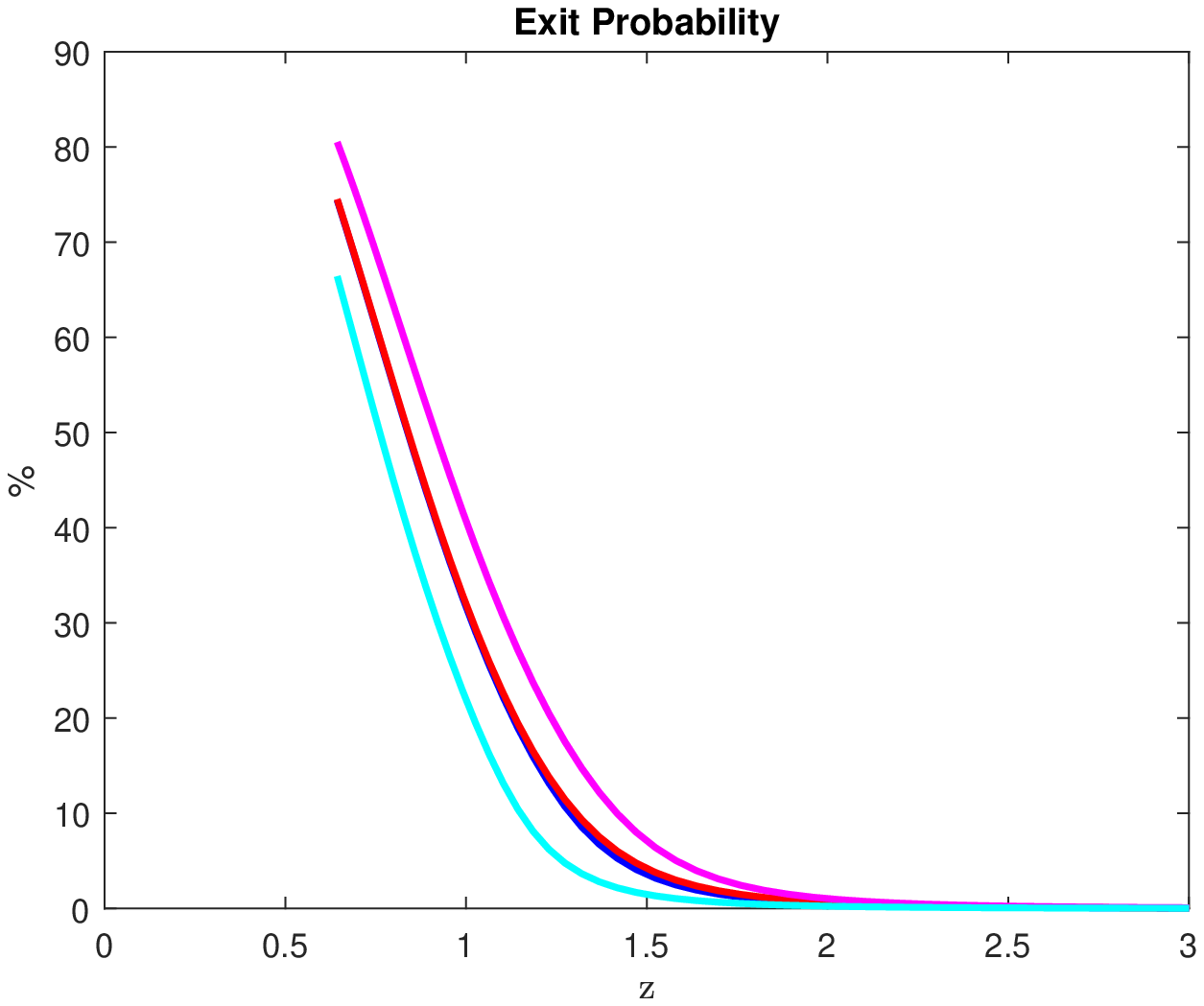}\\
    %\footnotesize \parbox[t]{0.65 in}{Notes:}\parbox[t]{5 in}{}
\end{figure}

To illustrate how the firms' entry and exit decisions depend on prices, Figure~\ref{fig:partialEqResponses} plots entry and exit probabilities in the stationary equilibrium as a function of idiosyncratic productivity, along with these probabilities when one price at a time is varied.\footnote{The entry probability (before observing the draw of the entry cost) is the probability that a potential entrant with productivity draw $z$ chooses to enter (i.e., $G_{e}(e_{t}^{*}(z))$); this probability is distinct from the entry \textit{rate}, which is the mass of actual entrants at each level of productivity divided by the measure of firms operating at that level of productivity. The exit probability is the probability (before observing the draw of the fixed operating cost) that an operating firm with productivity draw $z$ chooses to exit (i.e., $1-G_{c}(c_{t}^{*}(z))$); in the stationary equilibrium, this probability coincides with the exit \textit{rate}.} All else equal, a higher real interest rate ($r$) increases the exit probability across the productivity distribution, because firms place less weight on future profits when comparing their draw of the fixed operating cost against the expected value of continuing to operate. A higher real wage ($w$) directly decreases firm profits and thus decreases the value of continuing to operate, and so exit probabilities are higher. Conversely, a higher relative price of production firms' output ($p$) directly increases firm profits and yields lower exit probabilities. Similar reasoning applies to the profile of entry probabilities. Entry and exit probabilities tend to be more sensitive to changes in prices at lower levels of productivity than at higher levels, which reflects differences in the effect of price changes on the value of operating as well as the shape of the distribution of fixed costs.

\subsection{Dynamic Effects of a Monetary Policy Shock}
\label{subsec:dynamics}

I trace out the effects of a monetary policy shock in the calibrated model by solving for the model's equilibrium dynamics in the vicinity of the stationary equilibrium using a first-order perturbation solution method (see Appendix~\ref{sec:equilibriumconditions} for details). I assume that the parameter in the AR(1) process for the monetary policy shock ($\rho_{m}$) is 0.5, which is the same value used in \textcite{Ottonello_Winberry_2020}. I qualitatively compare the responses of entry and exit rates in the heterogeneous-firm (HF) model against those obtained using the proxy SVAR in Section~\ref{sec:empirical}, and explore the role of firm dynamics in transmitting the shock by comparing impulse responses against those from a comparable representative-firm (RF) model. The RF model replaces the mass of heterogeneous production firms with a representative firm that chooses labour input to maximise profits subject to a production function with decreasing returns to scale. The firm operates in competitive output and labour input markets and does not pay a fixed operating cost. In Appendix~\ref{sec:RANK}, I show that this RF model is identical to the standard three-equation New Keynesian model, except that decreasing returns to scale in production induce a steeper Phillips Curve than under a linear production technology. The parameter values in the calibrated RF model are the same as those in the HF model.

\begin{figure}[p]
    \center
    \caption{Impulse Responses to Monetary Policy Shock in HF and RF Models}
    \vspace{5pt}
    \label{fig:IRFs_HANKvsRANK}
    \includegraphics[scale=0.9]{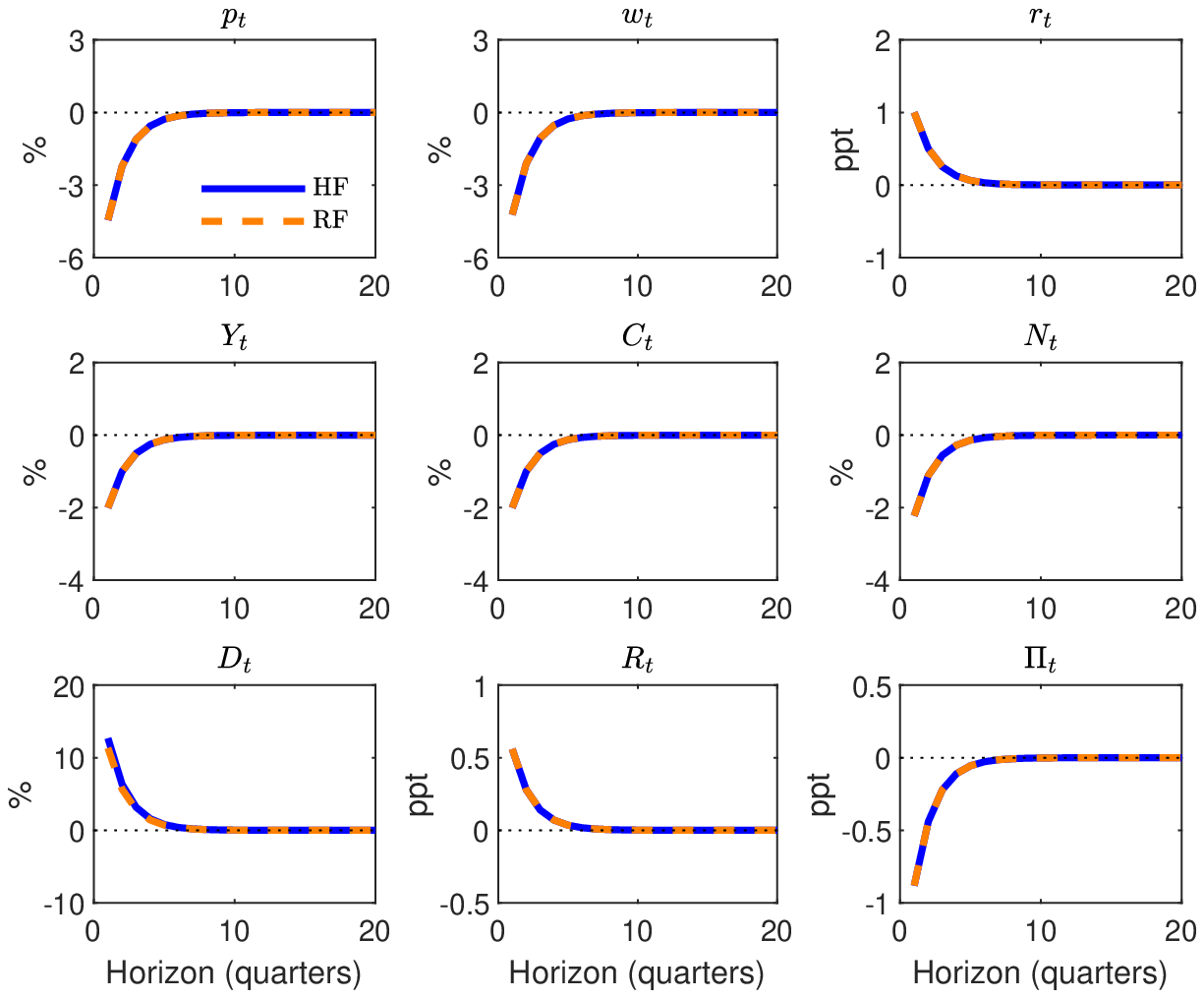} \\
    \vspace{10pt}
    \includegraphics[scale=0.85]{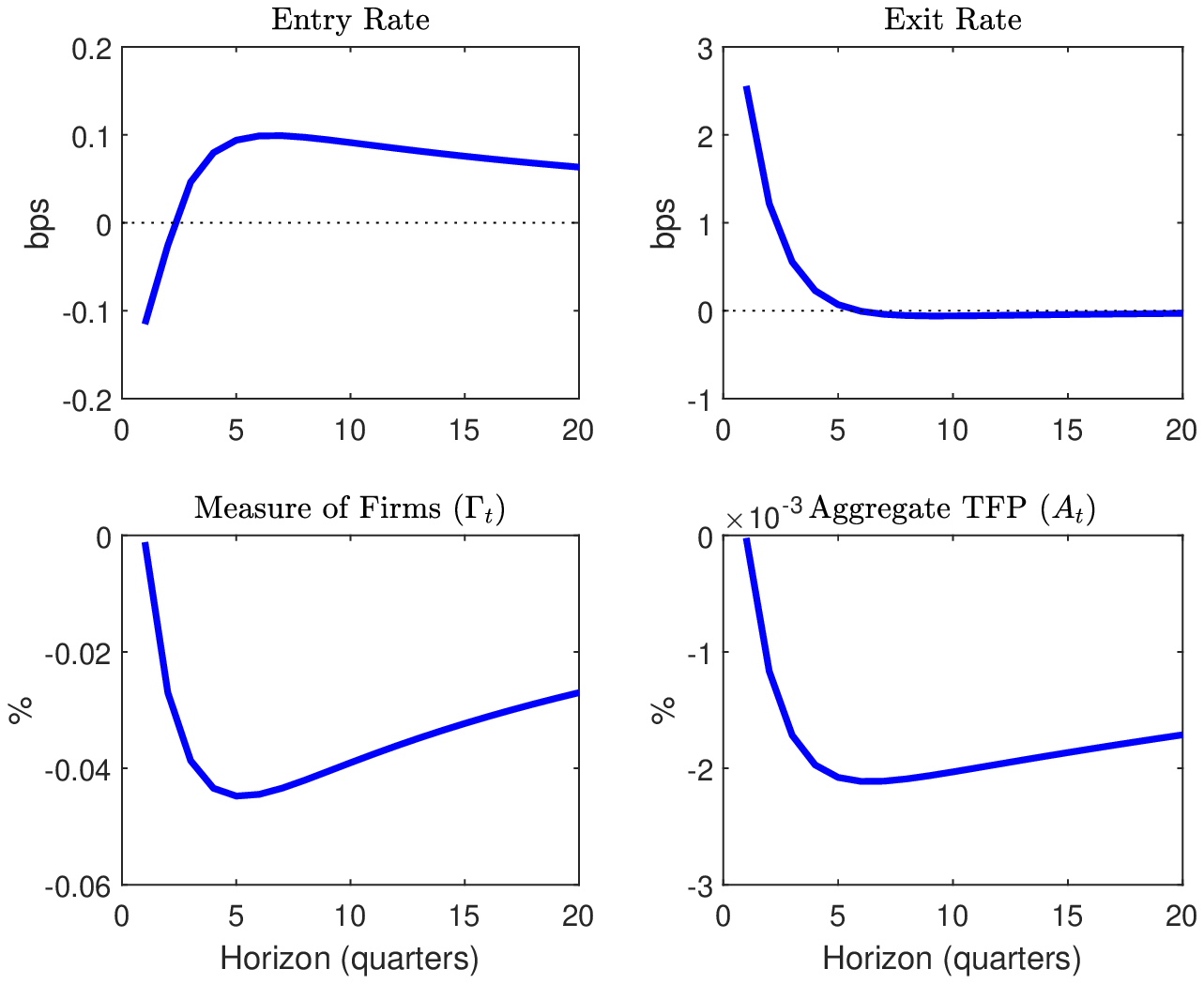}
    \footnotesize \parbox[t]{0.65 in}{Notes:}\parbox[t]{5 in}{Impulse responses are scaled so that the real interest rate increases by 1~percentage point on impact.}
\end{figure}

\bigskip

\noindent\textbf{Summary of responses.} Figure~\ref{fig:IRFs_HANKvsRANK} plots impulse responses to a positive monetary policy shock scaled so that the \textit{ex ante} real interest rate increases by one~percentage point on impact. Consumption, output and employment decline in response to the shock, as does the relative price of production firms' output and the real wage. The exit rate increases by about 2.5~basis points on impact before declining towards its stationary-equilibrium value.\footnote{The definitions of entry and exit rates in this exercise are consistent with measurement in the BED. The exit rate in period $t$ is equal to the measure of firms who choose to exit at the end of period $t$ divided by the measure of operating firms averaged over periods $t$ and $t+1$: $\int  (1-G_{c}(c_{t}^{*}(z)))d\mu_{t}(z)/X_{t+1}$, where $X_{t+1} = 0.5(\Gamma_{t+1}+\Gamma_{t})$ and $\Gamma_{t} = \int d\mu_{t}(z)$ is the total measure of operating firms. The entry rate in period $t$ is the measure of potential entrants who choose to enter in period $t$ divided by the measure of operating firms averaged over periods $t$ and $t-1$: $M\int G_{e}(e_{t}^{*}(z))dQ(z)/X_{t}$.} The entry rate initially declines by about 0.1~basis points, but then increases to be higher than in the stationary equilibrium, which reflects a decline in the denominator due to the increase in the exit rate. The increase in the exit rate, the decrease in the entry rate and the larger response of the exit rate relative to the entry rate are qualitatively consistent with the results from the proxy SVAR presented in Section~\ref{sec:empirical}, although the responses are clearly quantitatively small. The changes in firm entry and exit induce a small but persistent decrease in the total measure of operating firms ($\Gamma_t$), which is still around 0.03~per cent lower after five years.

Notably, the responses of macroeconomic variables to the shock are essentially indistinguishable across the HF and RF models; for example, output declines by 2.0023~per cent in the HF model compared with 2~per cent in the RF model. Allowing for the entry and exit of heterogeneous firms in an otherwise textbook New Keynesian model therefore has no substantive effect on the responses of most variables to a monetary policy shock.\footnote{An exception is the real dividend, whose increase on impact is about 1.4~percentage points larger in the HF model; this is partly due to a decline in fixed operating costs paid by production firms.} I discuss this result in more detail below.

\bigskip

\noindent\textbf{What drives the responses of entry and exit rates?} The entry probability at a particular level of idiosyncratic productivity is $G_{c}(e_{t}^{*}(z))$ and the exit probability is $1-G_{c}(c_{t}^{*}(z))$, where $e_t^{*}(z) = V_{t}(z)$ and $c_{t}^{*}(z) = \mathbb{E}_{t}(\Lambda_{t,t+1}V_{t+1}(z))$. The responses of these probabilities to the monetary policy shock will therefore depend on the shape of the fixed-cost distribution, the response of (expected) firm value to changes in the entire sequence of future prices, and the equilibrium changes in prices themselves. The responses of entry and exit rates then depend on the responses of entry and exit probabilities across the distribution of idiosyncratic productivity for potential entrants and operating firms, respectively.

Focusing on the response of the exit rate, the increase in the real interest rate is a decrease in the stochastic discount factor used by production firms to discount future profits. When deciding whether to pay the fixed operating cost to continue operating in the future, the production firms place less value on continuing to operate. Consequently, a larger proportion of firms draw a fixed operating cost that exceeds their exit threshold and, all else equal, the exit rate increases. An increase in the real interest rate also decreases household demand for the final good, which -- due to sticky prices -- decreases intermediate goods firms' demand for the undifferentiated good. This drives down the relative price of production firms' output ($p_t$) and thus reduces the value to production firms of operating, which would tend to increase the exit rate. However, because production firms operate in competitive markets, any reduction in the relative price of their output must be met with a decline in real marginal costs. Given decreasing returns to scale in production, this is achieved through production firms reducing employment, which pushes down the real wage ($w_t$). All else equal, this decline in the real wage would tend to result in a decrease in the exit rate. Similarly, changes in $w_{t}$ due to changes in labour supply will be transmitted through to changes in $p_{t}$. This dynamic induces a strong positive co-movement between $p_{t}$ and $w_{t}$.

\begin{figure}[h]
    \center
    \caption{Contributions of Prices to Responses of Entry and Exit Rates}
    \label{fig:priceContributions}
    \includegraphics[scale=0.75]{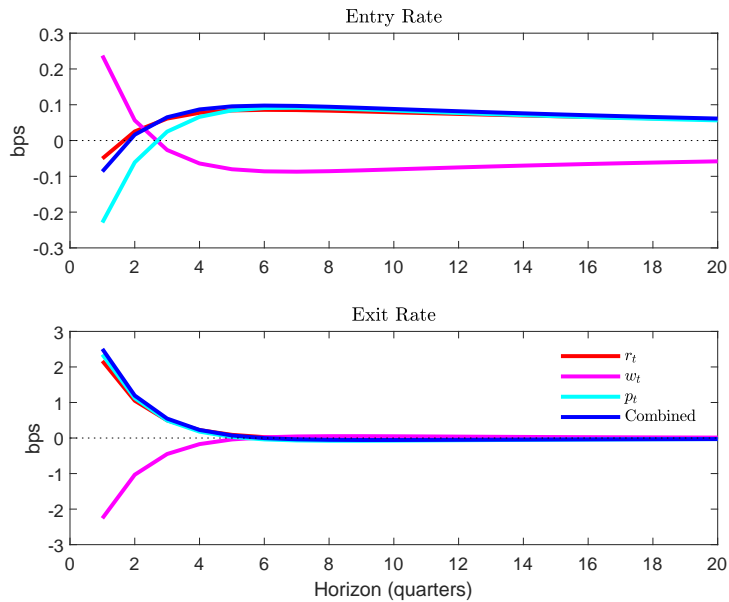} \\
    \footnotesize \parbox[t]{0.65 in}{Notes:}\parbox[t]{5 in}{Each line represents the response of entry or exit rates given the equilibrium path of a particular price following the shock, holding other prices constant and assuming that firms have perfect foresight about the path of prices; blue line is the overall response.}
\end{figure}

Figure~\ref{fig:priceContributions} plots the paths of the entry and exit rates obtained by feeding the equilibrium response of each price into the firms' problem while holding other prices constant and assuming that firms have perfect foresight about the path of prices. It is evident that the movements in $w_t$ and $p_t$ have effects on entry and exit rates that largely offset one another. Consequently, the responses of entry and exit rates are due largely to the direct effect of the change in the real interest rate.

\bigskip

\begin{figure}[h]
    \center
    \caption{Difference in Response of Labour Demand Relative to RF model}
    \vspace{5pt}
    \label{fig:employmentResponse}
    \includegraphics[scale=0.75]{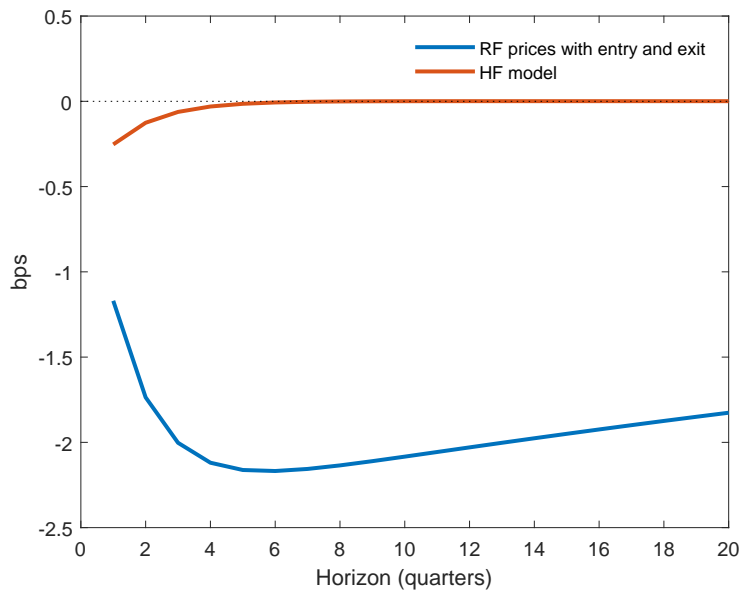} \\
    \footnotesize \parbox[t]{0.65 in}{Notes:}\parbox[t]{5 in}{Blue line is the response of labour demand to the path of prices obtained from the RF model (assuming perfect foresight about prices) minus the response of labour demand in the RF model; orange line is the difference between the responses of labour demand in the HF and RF models.}
\end{figure}

\noindent\textbf{Amplification and propagation of monetary policy shocks.} In the RF model, a reduction in $p_t$ results in the firm reducing its scale of operation, which reduces labour demand and thus puts downward pressure on the real wage. This intensive-margin adjustment also occurs in the HF model, with the degree of adjustment for each firm depending on the level of idiosyncratic productivity, but there is an additional extensive-margin adjustment; a persistent reduction in $p_t$ reduces the value of continuing, which reduces the threshold value of the fixed operating cost above which firms continue operating, so a greater proportion of firms exit at each level of productivity. Similarly, a reduction in $p_t$ results in a lower proportion of potential entrants entering at each level of productivity. These extensive-margin adjustments result in an additional decline in labour demand. To illustrate, I feed in the paths of prices from the RF model into the decision problem of production firms in the HF model (assuming perfect foresight about prices). This generates a fall in employment that is larger than the fall in employment in the RF model (Figure~\ref{fig:employmentResponse}, blue line). In the HF model, this additional margin of adjustment puts further downward pressure on the real wage. As discussed above, a reduction in the real wage is a reduction in real marginal cost for production firms, which must be met with a reduction in $p_t$ due to competition in the market for the production good. The entry and exit margins therefore amplify the effects of the monetary policy shock on $w_t$ and $p_t$ relative to the responses in the RF model. In general equilibrium, the overall effect of the larger responses of $w_t$ and $p_t$ in the HF model is to dampen the decline in $N_t$ that is directly due to the entry and exit margins (Figure~\ref{fig:employmentResponse}, orange line). $p_t$ is the real marginal cost faced by intermediate goods producers, so a larger decline in $p_t$ will also be associated with a larger decline in inflation.

Why is there little additional endogenous amplification of the monetary policy shock due to the entry and exit margins in this model? This partly reflects the small responses of entry and exit rates to the shock. However, I show below that there is still little amplification in a version of the model that has been modified to target the responses of entry and exit rates estimated using the proxy SVAR in Section~\ref{sec:empirical}. The other reason for the small degree of amplification is that the firms that are induced to exit (or not enter) due to the shock tend to be lower-productivity firms, which are small. The entry and exit margins therefore contribute only a small additional reduction in labour demand above and beyond the reduction in labour demand due to firms operating at smaller scale. Additionally, as described above and illustrated in Figure~\ref{fig:employmentResponse}, the role of entry and exit in amplifying the shock's effects on aggregate employment -- which is quantitatively small to begin with -- is dampened by the general-equilibrium responses of prices.

The entry and exit margins also serve to endogenously propagate the monetary policy shock. The state variable in the RF model is the monetary policy shock, so there is no endogenous state and no endogenous propagation of shocks (the $p$th-order autocorrelation of any variable in the RF model is simply $\rho_m^p$). In contrast, the measure of incumbent firms over idiosyncratic productivity is the endogenous state in the HF model, which is shaped by changes in entry and exit rates across the productivity distribution, so there is endogenous propagation of the monetary policy shock. However, given the similarity of responses in the HF and RF models, this endogenous propagation mechanism appears to be quantitatively unimportant. For example, the four-quarter autocorrelation of output is 0.064 in the HF model compared with 0.0625 in the RF model.

\bigskip

\noindent\textbf{The endogenous response of aggregate TFP.} Aggregating individual production firms' output yields an aggregate production function of the form $Y_{t} = A_{t}N_{t}^{\nu}$, where
\begin{equation} \label{eq:aggregateTFP}
    A_{t} = \left[\Gamma_{t}\mathbb{E}_{z}\left(z^{\frac{1}{1-\nu}}\right)\right]^{1-\nu}
\end{equation}
and $\mathbb{E}_{z}(.)$ is with respect to the distribution of firms, $\mu_{t}(z)/\Gamma_{t}$. The variable $A_{t}$ is a natural measure of aggregate TFP. Changes in entry and exit rates due to a monetary policy shock induce changes in the measure of firms operating. If the responses of entry and exit rates differ across the distribution of firms, the distribution over productivity will also respond to the shock. Aggregate TFP in this model is therefore endogenous, which is a channel of monetary policy transmission that is absent in the textbook New Keynesian model.

\begin{figure}[h]
    \center
    \caption{Change in Distribution of Firms}
    \vspace{5pt}
    \label{fig:dMu}
    \includegraphics[scale=0.75]{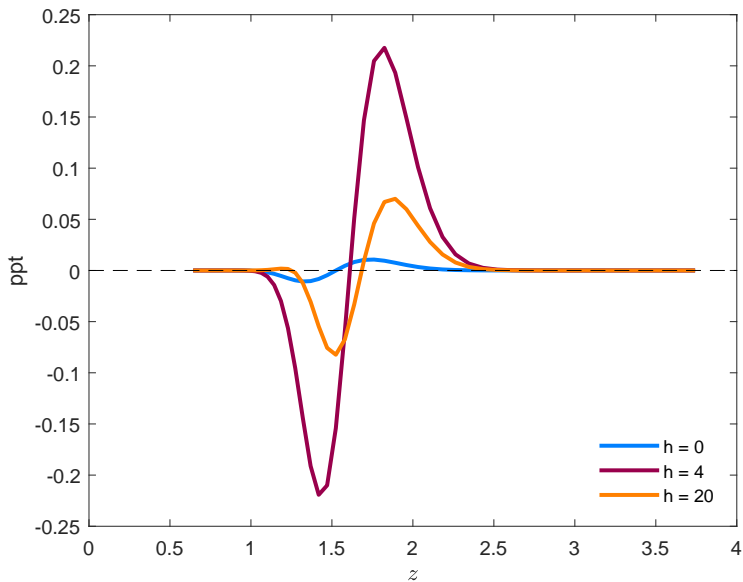} \\
    \footnotesize \parbox[t]{0.65 in}{Notes:}\parbox[t]{5 in}{$h$ is the number of quarters after the monetary policy shock; distribution of firms is $\mu_{t}(z)/\Gamma_{t}$.}
\end{figure}

Figure~\ref{fig:IRFs_HANKvsRANK} shows that the shock causes a small, but highly persistent, decrease in aggregate TFP. Figure~\ref{fig:dMu} plots the change in the distribution of firms at each level of productivity for different horizons after the realisation of the monetary policy shock. On impact, there is a decline in the measure of firms operating at all levels of productivity due to a decline in entry rates (operating firms who choose to exit following the shock only do so in the subsequent period). This decline tends to be larger for firms with lower productivity, so there is a small shift upwards in the productivity distribution, which increases one of the two components of aggregate TFP, $\mathbb{E}_{z}\left(z^\frac{1}{1-\nu}\right)$. However, the decline in $\Gamma_{t}$ more than offsets this effect, so aggregate TFP falls slightly. Four quarters after the shock, the change in the distribution of firms over idiosyncratic productivity is more pronounced due to the increase in exit rates, which again tends to be larger for lower-productivity firms, but the decline in the measure of firms operating continues to outweigh this effect, so aggregate TFP is still lower. Subsequently, the distribution of firms only slowly reverts to its stationary distribution due to the persistent process for idiosyncratic productivity. The change in entry and exit rates has long-lasting effects on both the productivity distribution and the total measure of operating firms, so the change in aggregate TFP is extremely persistent (the four-quarter autocorrelation in $A_t$ is around 0.95), albeit small in magnitude.\footnote{Consistent with the model's prediction, \textcite{Moran_Queralto_2018} estimate that aggregate TFP declines following a contractionary monetary policy shock using an SVAR in which the monetary policy shock is identified using a causal ordering. Adding measures of aggregate TFP from \textcite{Fernald_2012} to the proxy SVAR from Section~\ref{sec:empirical} also suggests that aggregate TFP may decline following a contractionary shock, although the 90~per cent confidence intervals for the response of utilisation-adjusted TFP include zero at all horizons.}

\bigskip

\noindent\textbf{What have we learned?} Even though the monetary policy shock is endogenously amplified and propagated by changes in entry and exit rates and a resulting change in aggregate TFP, these channels of transmission appear to be quantitatively unimportant in the sense that impulse responses from a comparable model without these channels are almost identical. A pertinent question is the extent to which this result depends on the calibration or on particular features of the model's structure. I explore this in the next section.

\subsection{Robustness}
\label{subsec:robustness}

\begin{figure}[h]
    \center
    \caption{Responses Under Alternative Values of $\nu$}
    \label{fig:nu}
    \includegraphics[scale=1]{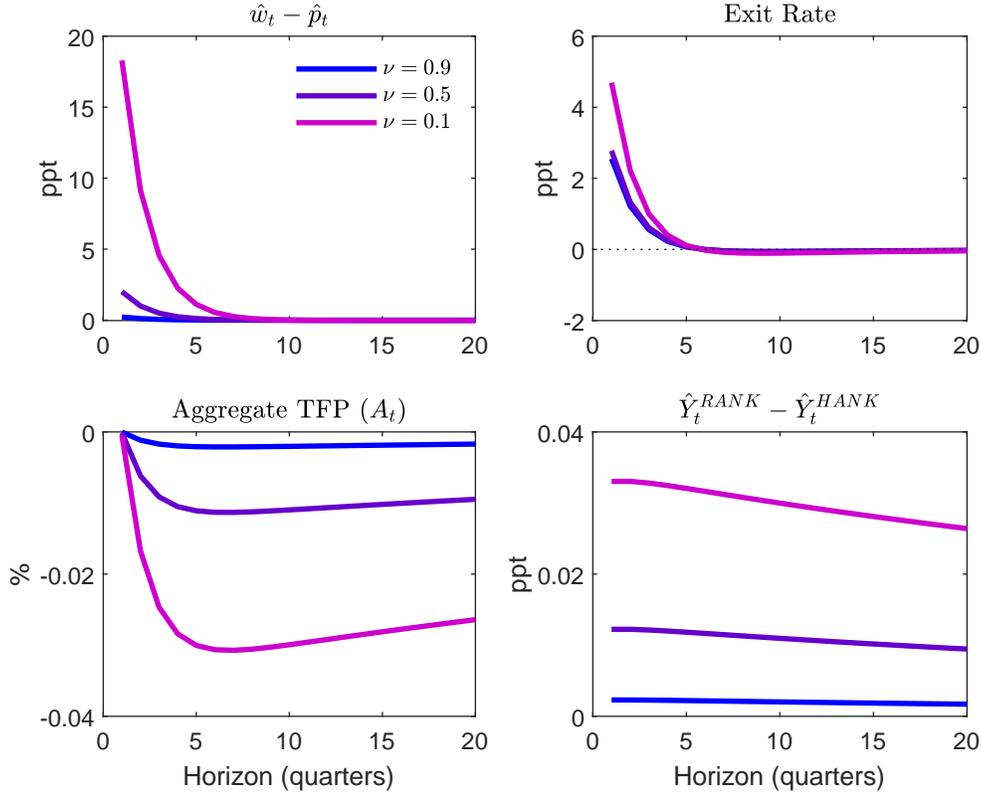} \\
    \footnotesize \parbox[t]{0.65 in}{Notes:}\parbox[t]{5 in}{Under each value of $\nu$, model is recalibrated to target the same moments as in Section~\ref{subsec:calibration}; $\hat{x}_{t}$ is the log-deviation of the variable $x_t$ from its value in the stationary equilibrium.}
\end{figure}

\noindent\textbf{Returns to scale and the co-movement of $w_t$ and $p_t$.} As discussed above, a key feature of the model's responses to the shock is the strong co-movement of $w_t$ and $p_t$, which results in the responses of entry and exit rates being predominately driven by the direct effect of the change in the real interest rate. The strength of this co-movement will be affected by the degree of returns to scale, which is controlled by $\nu$. With lower $\nu$, a given decline in $p_{t}$ will result in a smaller decline in labour demand than when $\nu$ is higher, because more-strongly decreasing returns to scale mean that a smaller reduction in employment is required to achieve the same reduction in marginal cost. In turn, the smaller decline in labour demand yields a smaller decline in $w_{t}$. A smaller decline in $w_t$ for any decline in $p_t$ means that entry and exit rates will respond more strongly to the shock. Smaller values of $\nu$ therefore generate larger changes in entry and exit rates, and larger differences between the responses of variables in the HF and RF models. However, even implausibly small values of $\nu$ do not generate substantive differences between the responses of variables in the HF and RF models. For example, assuming $\nu = 0.1$ generates a large wedge between the responses of $p_t$ and $w_t$, but the exit rate increases by only 4--5~basis points and the decline in output is only about 0.03~percentage points larger in the HF model than in the RF model (Figure~\ref{fig:nu}).

\bigskip

\noindent\textbf{Sensitivity to other parameter values.} Changing the parameters of the model one-at-a-time without recalibrating the model tends to have little impact on the qualitative results with the exception of the parameters characterising the distribution of fixed costs. This is because the responses of entry and exit rates to the shock depend partly on the shape of the distribution of fixed costs. For example, increasing $\mu_c$ or $\sigma_c$ sufficiently can yield responses of entry and exit rates to the shock that are similar to those estimated in the data. This generates larger differences between the responses of variables in the HF and RF models, although these difference are still small (e.g., the difference in the responses of output is on the order of one-tenth of a percentage point) and come at the cost of an unrealistically high exit rate in the stationary equilibrium.

\bigskip

\noindent\textbf{Matching the empirical responses of entry and exit rates.} The impact responses of entry and exit rates in the model are smaller in magnitude than the peak responses estimated in the data. A natural question is whether a model with empirically realistic responses of entry and exit rates can generate meaningfully different impulse responses to the RF model. One way to increase the responsiveness of entry and exit rates to the shock without substantively altering the structure of the model or changing its stationary equilibrium is to allow the distribution of fixed costs to depend directly on the deviation of the real interest rate from its value in the stationary equilibrium. In particular, I suppose that $c \sim LN(\mu_c + \alpha_{c}(r_{t} - 1/\beta),\sigma_{c})$ and $e \sim LN(\mu_c + \alpha_{e}(r_{t} - 1/\beta),\sigma_{c})$, where $\alpha_{c}$ and $\alpha_{e}$ are calibrated such that the exit rate increases by 10~basis points and the entry rate decreases by 4.5~basis points in response to the shock. Even with these larger responses of entry and exit rates, the responses of macroeconomic variables are still very similar in the HF and RF models; for example, the difference in the impact response of output is only about one-tenth of a percentage point.

\bigskip

\noindent\textbf{Alternative model structures.} The similarity between the responses of variables in the HF and RF models also holds under several alternative assumptions about the structure of the model. These include the assumptions that: fixed costs are denominated in either units of the production good or labour (rather than the final good); potential entrants who pay the entry cost do not operate until the subsequent period; production firms are risk neutral; or there is free entry of firms instead of a fixed mass of potential entrants. It also holds in a version of the model that adds physical capital and other features of typical medium-scale New Keynesian models, including aggregate investment-adjustment costs, habits in consumption and sticky wages. These alterations of the model are described in more detail in Appendix~\ref{sec:alternativespecifications}.

The reason why the responses of variables in the HF and RF models remain similar under these alternative specifications is that movements in $p_t$ and the other prices that enter the production firms' decision problem continue to largely offset one another, so the responses of entry and exit rates are driven predominately by the direct effect of the real interest rate. Consequently, the responses remain quantitatively small. Moreover, the firms that are induced to exit (or not enter) still tend to be those with lower productivity (i.e., small firms), so responses along the extensive margin do not result in a large additional reduction in labour demand.

\bigskip

\noindent\textbf{Firm-level frictions?} As discussed above, part of the reason for the limited additional amplification of the monetary policy shock in the HF model relative to the RF model is that the firms that are induced to exit (or not enter) due to the shock tend to be low-productivity firms, which are small. Changes in entry and exit therefore result in little additional reduction in labour demand above and beyond the reduction in labour demand due to changes along the intensive margin, and there is little additional effect on equilibrium prices or aggregate quantities. Frictions that make larger firms' entry and exit decisions more sensitive to changes in prices should result in these margins playing a more important role in the transmission of monetary policy shocks, because any given change in entry or exit rates will result in a larger change in labour demand, and thus greater amplification of the shock. For example, the presence of labour-adjustment costs (as in \textcite{Hopenhayn_Rogerson_1993}) would reduce the ability of operating firms to adjust employment along the intensive margin in response to changes in prices and could result in more exit among larger firms. Models with investment-adjustment costs (as in \textcite{Clementi_Palazzo_2016}) or financial frictions (as in \textcite{Ottonello_Winberry_2020}) may also be able to generate more-substantive roles for entry and exit in amplifying and propagating monetary policy shocks if these frictions make the entry and exit decisions of larger firms (in terms of employment or capital) more sensitive to changes in prices. I leave exploration of these avenues for future work.\footnote{In assessing the role of entry and exit in amplifying the shock in a model with firm-level frictions, one would want to compare responses against comparable models. The relevant model may be a representative-firm model with the given friction, or a heterogeneous-firm model with the given friction but without endogenous entry and exit. A version of my model without endogenous entry and exit would be equivalent to a representative-firm model, since $\mu_{t}(z)$ would be constant.}

\section{Conclusion}
\label{sec:conclusion}

A New Keynesian extension of the standard firm-dynamics model is able to generate qualitatively similar responses of firm entry and exit rates to a monetary policy shock as those estimated using a proxy SVAR, in the sense that firms are more likely to exit and less likely to enter following a contractionary shock. In the model, the decrease in entry and the increase in exit tend to be larger for lower-productivity firms, which shifts the distribution of idiosyncratic TFP upwards. A decline in the measure of operating firms due to the decrease in entry and the increase in exit more than offsets the upward shift in the productivity distribution and results in aggregate TFP falling. However, the decline in aggregate TFP is quantitatively small and the responses of macroeconomic variables in the model are essentially indistinguishable from those in a comparable representative-firm model. Accounting for firm dynamics therefore appears to have little effect on predictions about the effects of monetary policy relative to those from a representative-firm model. This also holds under different parameter values and under alternative assumptions about the structure of the model.

I have deliberately abstracted from firm-level frictions to highlight the role (or lack thereof) of firm dynamics in the transmission of monetary policy in a simple New Keynesian firm-dynamics model. Future work could explore the role of entry and exit in monetary policy transmission using models with firm-level frictions. Frictions that result in larger firms' entry and exit behaviour being more sensitive to changes in prices are likely to result in these margins playing a more important role in the transmission of monetary policy shocks. Firm-level evidence on the entry and exit behaviour of firms following a monetary policy shock may be useful in disciplining the responses of entry and exit across the distribution of firms and in assessing the relevance of alternative frictions.

%\section{Acknowledgements}
%I thank Morten Ravn, Vincent Sterk and seminar participants at UCL for useful feedback. The views in this paper do not reflect those of the Reserve Bank of Australia, whose financial support I gratefully acknowledge. This research did not receive any specific grant from funding agencies in the public, commercial, or not-for-profit sectors.

\newpage
\printbibliography
\addcontentsline{toc}{section}{References}

\newpage
\numberwithin{equation}{section}
\setcounter{equation}{0}
\begin{appendices}
    \section{Equilibrium Conditions and Solution Method}
\label{sec:equilibriumconditions}

\noindent\textbf{Intermediate goods producers.} There is a unit mass of intermediate goods producers indexed by $i \in [0,1]$. These firms purchase the undifferentiated good from the production firms and use it to produce a differentiated good, denoted $\tilde{Y}_{it}$, according to the production function $\tilde{Y}_{it} = Y_{it}$, where $Y_{it}$ is the quantity of the undifferentiated good demanded by firm $i$. The intermediate goods producers sell their products to the representative final good producer in a monopolistically competitive market, taking the demand schedule of the final good producer (derived below) as given. Intermediate goods producers face a quadratic price-adjustment cost -- paid in units of the final good -- as in \textcite{Rotemberg_1982}. These firms discount the future using the stochastic discount factor of the representative household.

The profit-maximisation problem faced by intermediate goods producers is
\begin{equation}
  \max_{\{P_{it},Y_{it},\tilde{Y}_{it}\}_{t=0}^{\infty}} \mathbb{E}_{0}\sum_{t=0}^{\infty}\Lambda_{0,t}\left\{\frac{P_{it}}{P_t}\tilde{Y}_{it}-p_{t}Y_{it} - \frac{\xi}{2}\left(\frac{P_{it}-P_{i,t-1}}{P_{i,t-1}}\right)^{2}Y_{t}\right\}
\end{equation}
subject to
\begin{align}
  \tilde{Y}_{it} &= Y_{it} \\
  \tilde{Y}_{it} &= \left(\frac{P_{it}}{P_{t}}\right)^{-\gamma}Y_{t},
\end{align}
where $P_{it}$ is the price of the intermediate good sold by firm $i$, $Y_{t}$ is aggregate production of the final good and $\gamma$ is the elasticity of substitution between differentiated goods in final good production. The first-order condition for the problem is
\begin{multline}
  \left\{(1-\gamma)\left(\frac{P_{it}}{P_{t}}\right)^{-\gamma} + \gamma p_{t}\left(\frac{P_{it}}{P_{t}}\right)^{-\gamma-1} - \xi\left(\frac{P_{it}}{P_{i,t-1}}-1\right)\frac{P_{t}}{P_{i,t-1}}\right\}\frac{Y_{t}}{P_{t}}  \\
  +\xi\mathbb{E}_{t}\left[\Lambda_{t,t+1}\left(\frac{P_{i,t+1}}{P_{it}}-1\right)\frac{P_{i,t+1}}{P_{it}^{2}}Y_{t+1}\right] = 0.
\end{multline}
I consider symmetric equilibria where all intermediate goods producers face the same initial price, which implies that they will optimally choose the same price in each period (i.e. $P_{it} = P_{t}$). This assumption implies that the first-order condition above simplifies to the nonlinear Phillips Curve in Equation~\ref{eq:pricesetting}.

\bigskip

\noindent\textbf{Final good producer.} A representative final good producer purchases differentiated goods from the intermediate goods producers and bundles them into a final good according to the production function
\begin{equation}
  Y_{t} = \left(\int_{0}^{1}\tilde{Y}_{it}^{\frac{\gamma-1}{\gamma}}di\right)^{\frac{\gamma}{\gamma-1}}. \label{eq:finalgoodprodfn}
\end{equation}
It sells the final good to households (for consumption) and intermediate goods producers (to pay price-adjustment costs) in a competitive market at price $P_{t}$. The profit-maximisation problem of the final good producer is
\begin{equation}
  \max_{\{Y_{t},\{\tilde{Y}_{it}\}_{i \in [0,1]}\}} P_{t}Y_{t} - \int_{0}^{1}P_{it}\tilde{Y}_{it}di
\end{equation}
subject to Equation~\ref{eq:finalgoodprodfn}. The first-order conditions for the problem yield the demand schedule
\begin{equation}\label{eq:finalgoodsfirmdemandschedule}
  \tilde{Y}_{it} = \left(\frac{P_{it}}{P_{t}}\right)^{-\gamma}Y_{t},
\end{equation}
where
\begin{equation}\label{eq:finalgoodprice}
  P_{t} = \left(\int_{0}^{1}P_{it}^{1-\gamma}di\right)^{\frac{1}{1-\gamma}}
\end{equation}
is the price index for the final good. Final goods producers make zero profit in equilibrium.

\bigskip

\noindent\textbf{Solution method.} Given the discretisation of the state-space for idiosyncratic productivity over a $k$-point grid, the production firms' Bellman equation and the transition function for $\mu_{t}(z)$ become a system of $2k$~equations. Denote the grid points for $z$ as $z_{i}$, $i=1,\ldots,k$, let $P_{ij} = \mathrm{Pr}(z' = z_{j}|z = z_{i})$ be the transition probabilities of the discrete-state Markov process induced by discretisation of the AR(1) process for $z$ and let $q(z_{i})$ be the probability that a potential entrant draws $z = z_{i}$ given the discretisation of $Q(z)$. Combining these equations with the other equilibrium conditions derived in Section~\ref{sec:model} yields the following system of equations:\footnote{I omit the definitions of $D_t$ and $T_t$ since they are not necessary to solve for the equilibrium dynamics of the other variables.}
\begin{multline}
  V_{t}(z_{i}) = p_{t}z_{i}\left(\frac{w_{t}}{\nu p_{t} z_{i}}\right)^{\frac{\nu}{\nu-1}} - w_{t}\left(\frac{w_{t}}{\nu p_{t} z_{i}}\right)^{\frac{1}{\nu-1}}
  + \Big[\sum_{j=1}^{k}P_{ij}\mathbb{E}_{t}(\Lambda_{t,t+1}V_{t+1}(z_{j})) \\
  - \mathbb{E}_{c}\left(c | c \leq \sum_{j=1}^{k}P_{ij}\mathbb{E}_{t}(\Lambda_{t,t+1}V_{t+1}(z_{j}))\right)\Big]G_{c}\left(\sum_{j=1}^{k}P_{ij}\mathbb{E}_{t}(\Lambda_{t,t+1}V_{t+1}(z_{j}))\right)
\end{multline}
\begin{equation}
    \mu_{t+1}(z_{i}) = \sum_{j=1}^{k}G_{c}\left(\sum_{k=1}^{k}P_{jk}\mathbb{E}_{t}(\Lambda_{t,t+1}V_{t+1}(z_{k}))\right)P_{ji}\mu_{t}(z_{j}) + M G_{e}(V_{t+1}(z_{i}))q(z_{i})
\end{equation}
\begin{equation}
  N_{t} = \sum_{i=1}^{k}\left(\frac{w_{t}}{\nu p_{t} z_{i}}\right)^{\frac{1}{\nu-1}}\mu_{t}(z_{i})
\end{equation}
\begin{equation}
    w_{t} = \kappa_{0}C_{t}^{\sigma}N_{t}^{\kappa_{1}}
\end{equation}
\begin{equation}
  1 = \mathbb{E}_{t}\left(\Lambda_{t,t+1}\frac{R_{t}}{\Pi_{t+1}}\right)
\end{equation}
\begin{equation}
  \frac{R_{t}}{R} = \left(\frac{\Pi_{t}}{\Pi}\right)^{\phi}\exp(\varepsilon_{t}^{m})
\end{equation}
\begin{equation}
  (1-\gamma) + \gamma p_{t} - \xi(\Pi_{t} - 1)\Pi_{t} = -\xi \mathbb{E}_{t}\left[\Lambda_{t,t+1}(\Pi_{t+1}-1)\Pi_{t+1}\frac{Y_{t+1}}{Y_{t}}\right]
\end{equation}
\begin{equation}
  Y_{t} = \sum_{i=1}^{k}\left(\frac{w_{t}}{\nu p_{t} z_{i}}\right)^{\frac{\nu}{\nu-1}}\mu_{t}(z_{i})
\end{equation}
\begin{equation}
    Y_{t} = C_{t} + \frac{\xi}{2}(\Pi_{t}-1)^{2}Y_{t}.
\end{equation}
%\begin{multline}
%    Y_{t} = C_{t} + \sum_{i=1}^{k}G_{c}\left(\sum_{j=1}^{k}P_{ij}\mathbb{E}_{t}(\Lambda_{t,t+1}V_{t+1}(z_{j}))\right)\mathbb{E}_{t}\left(c|c \leq \sum_{j=1}^{k}P_{ij}\mathbb{E}_{t}(\Lambda_{t,t+1}V_{t+1}(z_{j}))\right)\mu_{t}(z_{i}) \\
%    + M\lambda\sum_{i=1}^{k}G_{e}(V_{t}(z_{i}))\mathbb{E}_{c}(c | c \leq V_{t}(z_{i})/\lambda)q(z_{i}) + \frac{\xi}{2}(\Pi_{t}-1)^{2}Y_{t}
%\end{multline}
The first two equations must hold for $i=1,\ldots,k$. Given the definition of $\Lambda_{t,t+1}$ and treating $V_{t}(z_{i})$ and $\mu_{t}(z_{i})$ as endogenous variables, this is a system of $2k+7$ equations in $2k+7$ endogenous variables, which is differentiable in the variables. Consequently, I am able to solve the model using perturbation around the stationary equilibrium (see, for example, \textcite{Schmitt-Grohe_Uribe_2004}).\footnote{The Blanchard-Kahn conditions are satisfied at the calibrated parameter values, so the equilibrium is unique and stable. I have verified numerically that the Blanchard-Kahn conditions are satisfied only when $\phi > 1$ given the calibrated values of the other parameters, so the well-known `Taylor principle' holds in this model.} I use Dynare to solve the model using first-order perturbation (Adjemian \textit{et al.} 2020\nocite{Adjemian_etal_2020}).\footnote{The results are essentially identical when using second-order perturbation.}

\section{Representative-Firm Model}
\label{sec:RANK}

In this section, I describe the RF model used as a benchmark against which to compare the HF model developed in Section~\ref{sec:model}. The RF model replaces the mass of heterogeneous production firms in the HF model with a representative firm, who chooses labour input to maximise profits subject to a production function with decreasing returns to scale. The firm operates in competitive output and labour input markets. It does not pay a fixed operating cost, and there is no entry and exit. In real terms, the firm's (static) profit-maximisation problem is
\begin{equation}
    \max_{n_{t}}p_{t}n_{t}^{\nu} - w_{t}n_{t},
\end{equation}
with first-order condition $w_{t} = \nu p_{t}n_{t}^{\nu - 1}$.

The decision problems faced by the representative household, intermediate goods firms and the final goods firm are identical to those in the HF model, as is the central bank's policy rule. %After enforcing bond market clearing ($B_{t} = 0$), labour market clearing ($N_{t} = n_{t}$) and goods market clearing, the conditions characterising a symmetric equilibrium are:
%\begin{equation*}
%    w_{t} = \kappa_{0}C_{t}^{\sigma}N_{t}^{\kappa_{1}}
%\end{equation*}
%\begin{equation}
%  1 = \mathbb{E}_{t}\left(\Lambda_{t,t+1}\frac{R_{t}}{\Pi_{t+1}}\right)
%\end{equation}
%\begin{equation}
%  \frac{R_{t}}{R} = \left(\frac{\Pi_{t}}{\Pi}\right)^{\phi}\exp(\varepsilon_{t}^{m}),
%\end{equation}
%\begin{equation}
%    w_{t} = \nu  p_{t}N_{t}^{\nu - 1}.
%\end{equation}
%\begin{equation}
%    Y_{t} = N_{t}^{\nu}
%\end{equation}
%\begin{equation}
%  (1-\gamma) + \gamma p_{t} - \xi(\Pi_{t} - 1)\Pi_{t} = -\xi \mathbb{E}_{t}\left[\Lambda_{t,t+1}(\Pi_{t+1}-1)\Pi_{t+1}\frac{Y_{t+1}}{Y_{t}}\right]
%\end{equation}
%\begin{equation}
%    Y_{t} = C_{t} + \frac{\xi}{2}(\Pi_{t}-1)^{2}Y_{t}.
%\end{equation}
Log-linearising around a deterministic steady state with zero net inflation yields a system of three equations in three endogenous variables plus the exogenous monetary policy shock:
\begin{equation}
    \hat{Y}_{t} - \mathbb{E}_{t}\hat{Y}_{t+1} = -\frac{1}{\sigma}(\hat{R}_{t} - \mathbb{E}_{t}\hat{\Pi}_{t+1})
\end{equation}
\begin{equation}
    \hat{\Pi}_{t} = \frac{\gamma-1}{\xi}\left(\sigma + \frac{\kappa_{1}+1}{\nu} - 1\right)\hat{Y}_{t} + \beta\mathbb{E}_{t}\hat{\Pi}_{t+1}
\end{equation}
\begin{equation}
    \hat{R}_{t} = \phi\hat{\Pi}_{t} + \varepsilon_{t}^{m},
\end{equation}
where $\hat{x}_{t} = \ln x_{t} - \ln x$ is the log-deviation of the variable $x_{t}$ from its value in the deterministic steady state. The first equation is the Euler equation or IS curve, the second is the Phillips Curve and the last is the Taylor rule. Decreasing returns to scale in production ($\nu < 1$) imply that the Phillips Curve is steeper than in the textbook case with linear production technology.

\section{Alternative Assumptions About Model Structure}
\label{sec:alternativespecifications}

This section describes alternative assumptions about the structure of the model and briefly outlines how the results change under these alternative assumptions. Unless otherwise noted, each model is recalibrated to target the same moments as in Section~\ref{subsec:calibration}. Where appropriate, the RF model is modified to maintain comparability with the HF model.

\bigskip

\noindent\textbf{Units of fixed costs:} In the baseline model, entry and operating costs are paid in units of the final good. Alternative assumptions are that entry and operating costs are paid in units of labour or in units of the production good. Under either assumption, the entry rate increases and the exit rate decreases in response to the shock, because declines in $p_t$ and $w_t$ directly shift down the distribution of real fixed costs. This pattern of responses is inconsistent with the empirical evidence presented in Section~\ref{sec:empirical}. In contrast with the baseline model, where entry and exit amplify the effects of the shock, the responses of entry and exit rates in these versions of the model dampen the effects of the shock. The responses of macroeconomic variables remain essentially indistinguishable from those in the RF model.

\bigskip

\noindent\textbf{Delay between entry and production:} The baseline model assumes that entrants begin operating in the same period in which they pay the entry cost, whereas continuing firms pay the fixed operating cost in period $t$ to continue into period $t+1$. Assuming instead that entrants pay the entry cost in period $t$ to begin operating in period $t+1$ means that a potential entrant with idiosyncratic productivity $z$ and entry cost $e$ enters only if $e \leq \mathbb{E}_{t}(\Lambda_{t,t+1}V_{t+1}(z))$, so the real interest rate directly affects the entry decision. Excepting the delayed response of the entry rate, this alternative assumption does not substantively alter the results relative to the baseline model.

\bigskip

\noindent\textbf{Risk-neutral firms:} The baseline model assumes that production firms discount the future using the representative household's stochastic discount factor, $\Lambda_{t,t+1}$. Assuming instead that production firms are risk-neutral and discount the future at rate $\beta$ means that the real interest rate no longer directly affects the value of the firm. Under this assumption, the responses of entry and exit rates are substantially smaller than in the baseline model, so there is even less amplification and propagation of the shock due to these margins.

\bigskip

\noindent\textbf{Physical capital and aggregate frictions:} This version of the model adds physical capital and other features of typical medium-scale New Keynesian models, including aggregate investment-adjustment costs, consumption habits and sticky wages.

Assume that the representative household can, in addition to saving in the nominal bond, also invest in physical capital ($K_{t}$), which is rented to the production firms in a competitive spot market at real rental rate $r_{kt}$ and depreciates at rate $\delta$. Investment ($I_t$) is undertaken subject to an investment-adjustment cost such that capital evolves according to
\begin{equation}
  K_{t+1} = \left[1 - \frac{\tau}{2}\left(\ln \frac{I_{t}}{I_{t-1}}\right)^{2}\right]I_{t} + (1-\delta)K_{t}.
\end{equation}
Also, assume that there are habits in consumption, so the household maximises
\begin{equation}
  \mathbb{E}_{0}\sum_{t=0}^{\infty}\beta^{t}\left(\frac{\left(C_{t} - hC_{t-1}\right)^{1-\sigma}-1}{1-\sigma} - \kappa_{0} \frac{N_{t}^{1+\kappa_{1}}}{1+\kappa_{1}}\right).
\end{equation}
Further assume that the representative household competitively supplies undifferentiated labour services to monopolistically competitive unions, who differentiate this labour and sell it to competitive labour packers subject to a quadratic wage-adjustment cost. The household owns the unions and profits generated by these unions are remitted to the household as dividends. The labour packers bundle differentiated labour services to produce undifferentiated `final' labour services, which are sold to production firms in a competitive market.\footnote{This setup follows Bayer \textit{et al.} (2019)\nocite{Bayer_Born_Luetticke_2019}.}

Letting $\lambda_{t}$ and $\omega_{t}$ be the Lagrange multipliers on the budget constraint and the capital accumulation equation, respectively, the first-order conditions for the household's problem are:
\begin{equation}
  w_{t}\lambda_{t} = \kappa_{0}N_{t}
\end{equation}
\begin{equation}
  1 = \beta\mathbb{E}_{t}\left(\frac{\lambda_{t+1}}{\lambda_{t}}\frac{R_t}{\Pi_{t+1}}\right)
\end{equation}
\begin{equation}
  \omega_{t} = \beta\mathbb{E}_{t}\left[\omega_{t+1}(1-\delta) + \lambda_{t+1}r_{k,t+1}\right]
\end{equation}
\begin{equation}
  \lambda_{t}=\omega_{t}\left[1 - \frac{\tau}{2}\left(\ln \frac{I_{t}}{I_{t-1}}\right)^{2} - \tau\ln\frac{I_{t}}{I_{t-1}}\right] + \beta\tau\mathbb{E}_{t}\left[\omega_{t+1}\frac{I_{t+1}}{I_{t}}\ln\frac{I_{t+1}}{I_{t}}\right],
\end{equation}
where $\lambda_{t} = (C_{t}-hC_{t-1})^{-\sigma} - h\beta\mathbb{E}_{t}(C_{t+1}-hC_{t})^{-\sigma}$. The stochastic discount factor is now $\Lambda_{t,t+1} = \beta (\lambda_{t+1}/\lambda_t)^{-\sigma}$.

The structure for the labour market generates the following (nonlinear) wage Phillips Curve:
\begin{equation}
    1 - \zeta + \zeta \frac{w_{t}}{w_{t}^{F}} - \chi(\Pi_{t}^{w}-1)\Pi_{t}^{w} = -\chi\mathbb{E}_{t}\left[\Lambda_{t,t+1}(\Pi_{t+1}^{w}-1)\frac{\left(\Pi_{t+1}^{w}\right)^{2}}{\Pi_{t+1}}\frac{N_{t+1}}{N_{t}}\right],
\end{equation}
where $w_{t}$ is the real wage earned by households, $w_{t}^{F}$ is the real wage paid by firms and $\Pi_{t}^{w} = \Pi_{t}w_{t}^{F}/w_{t-1}^{F}$ is the gross rate of inflation in final wages. $w_{t}/w_{t}^{F}$ is real marginal cost for the labour packers, which is the inverse of the (gross) markup charged by labour unions.

Production firms operate a Cobb-Douglas production technology with decreasing returns to scale. They hire labour from the labour packers at wage rate $w_{t}^{F}$ and rent capital from the household at rental rate $r_{kt}$. The production firms' Bellman equation is
\begin{equation}\label{eq:bellman_existing1}
   V_{t}(z) = \max_{n\geq0,k\geq0}p_{t}z\left(n^{\theta}k^{1-\theta}\right)^{\nu} - w_{t}^{F}n -r_{kt}k + \int\max\left\{\mathbb{E}_{t}\left(\Lambda_{t,t+1}V_{t+1}(z')\right)-c,0\right\}dG_{c}(c).
\end{equation}
The first-order conditions for this problem imply closed-form policy functions for labour demand ($n_{t}(z)$) and capital demand ($k_{t}(z)$). Using these policy functions and the threshold for exit, the Bellman equation becomes
\begin{multline}
  V_{t}(z) = p_{t}z\left(n_{t}(z)^{\theta}k_{t}(z)^{1-\theta}\right)^{\nu} - w_{t}^{F}n_{t}(z) - r_{kt}k_{t}(z) \\
  +\left[\mathbb{E}_{t}\left(\Lambda_{t,t+1}V_{t+1}(z')\right)-\mathbb{E}_{c}\left(c|c < \mathbb{E}_{t}\left(\Lambda_{t,t+1}V_{t+1}(z')\right)\right)\right]G_{c}\left(\mathbb{E}_{t}\left(\Lambda_{t,t+1}V_{t+1}(z')\right)\right).
\end{multline}

Maintaining the assumption that fixed operating and entry costs are remitted lump-sum to households, clearing of the final goods market requires
\begin{equation}
    Y_{t} = C_{t} + I_t + \frac{\xi}{2}(\Pi_{t}-1)^{2}Y_{t},
\end{equation}
where $Y_t = \int z\left(n_{t}(z)^{\theta}k_{t}(z)^{1-\theta}\right)^{\nu}d\mu_{t}(z)$, clearing of the labour market requires
\begin{equation}
    N_t = \int n_t(z)d\mu_{t}(z) + \frac{\chi}{2}(\Pi_{t}^{w} - 1)^{2}N_{t}
\end{equation}
and clearing of the capital market requires $K_{t} = \int k_{t}(z)d\mu_{t}(z)$.

I recalibrate the model to target a labour share of $0.6$, which implies that $\theta = 0.8$ and I continue to set $\nu = 0.9$.\footnote{This implies that the capital share is 0.15, with the remainder of output going to profits.} I assume that $\delta = 0.025$, so $r_{k} = 1/\beta - 1 + \delta \approx 0.035$, and that $\tau = \xi$. In a stationary equilibrium with zero net inflation in prices and final wages, $w_{t}/w_{t}^{F} = (\zeta - 1)/\zeta$. I assume that the markup in wage-setting is the same as in price-setting, so $\zeta = \gamma$ and the slope of the wage Phillips Curve is the same as the price Phillips Curve, so $\chi = \xi$. I set the consumption habit parameter $h$ to 0.5.

Under these assumptions, the entry rate decreases, the exit rate increases and aggregate TFP decreases in response to a contractionary monetary policy shock. There are no substantive differences between the responses of macroeconomic variables in the HF model and a comparable RF model that includes the same frictions as the HF model. Despite the presence of sticky wages, the fact that capital is liquid and there are no labour-adjustment costs means that the structure for production still induces a strong positive co-movement between the prices that enter the production firms' problem. The movements in these prices have effects on entry and exit rates that largely offset each other, so that the responses of entry and exit rates continue to be primarily driven by the direct effect of the change in the real interest rate. Quantitatively small responses of entry and exit rates -- and the small size of the firms induced to exit (or not enter) -- result in limited additional amplification of the shock.

\bigskip

\noindent\textbf{Free entry:} Let $\tilde{M}_t$ be the mass of \textit{actual} entrants and assume that entrants must pay $\tilde{e}\exp(\alpha(\tilde{M}_t-\tilde{M}))$ (where $\tilde{e} >0$ and $\alpha > 0$ are parameters) before drawing $z$ from $Q(.)$. The assumption that the entry cost depends on the deviation of $M_t$ from its value in the stationary equilibrium is used to avoid an unrealistically large elasticity of the entry rate to monetary policy shocks under free entry. Once an entrant has paid the entry cost and drawn $z$, they face the same decision problem as an incumbent firm with productivity $z$.\footnote{A firm that pays the entry cost will always operate for at least one period, because the fixed operating cost does not need to be paid until the end of the period.} The value of a new entrant with idiosyncratic productivity $z$ is therefore $V_{t}(z)$ (given by Equation~\ref{eq:bellman_existing2}) and the expected value of an entrant before drawing $z$ is $V_{t}^{e} = \int V_{t}(z)dQ(z)$. Free entry requires that $V_{t}^{e} \leq \tilde{e}\exp(\alpha(M_t-M))$, with equality in any equilibrium with entry. I consider equilibria with entry, so the free-entry condition is $V_{t}^{e} = \tilde{e}\exp(\alpha(M_t-M))$.

The transition function for the measure of firms over idiosyncratic productivity under free entry is
\begin{equation}\label{eq:transitionmufreeentry}
  \mu_{t+1}(B) = \int\int_{z' \in B}G_{c}\left(c_{t}^{*}(z)\right)dF(z'|z)d\mu_{t}(z) + \tilde{M}_t\int_{z \in B}dQ(z).
\end{equation}
When calibrating this version of the model, as in the baseline model, I normalise the real wage to unity in the stationary equilibrium. I then solve for the parameter $\tilde{e}$ such that the free entry condition is satisfied. To solve for the mass of entrants in the stationary equilibrium, I first set $\tilde{M} = 1$ and iterate over Equation~\ref{eq:transitionmufreeentry} to convergence. As explained in \textcite{Hopenhayn_Rogerson_1993}, because $\mu(z)$ is homogeneous of degree one in $\tilde{M}$, this gives the stationary measure of firms up to scale. The equilibrium mass of entrants can then be solved for from the labour market clearing condition given the target for the employment-to-population ratio. Consumption is obtained from the household's budget constraint and $\kappa_{0}$ is backed out from the labour supply condition.

Assuming that $\alpha = 15$, the entry rate decreases by about 4.5~basis points, which is similar to the point estimate of its response in the proxy SVAR. The exit rate increases, but only by about 0.8~basis points. The larger response of the entry rate relative to the exit rate is qualitatively at odds with the point estimates from the proxy SVAR. There is a quantitatively small decline in aggregate TFP and the responses of variables in the HF and RF models are again extremely similar.

\end{appendices}

\end{document}